\documentclass{aa}
\usepackage{graphics}
\newcommand{\cmtwo}{cm$^{-2}$}
\newcommand{\cmthree}{cm$^{-3}$}

\newcommand{\pcq}{pc$^{2}$}

\newcommand{\kms}{km\,s$^{-1}$}       
\newcommand{\vs}{$v_{\rm s}$}
\newcommand{\no}{$n_{0}$}

\newcommand{\ecs}{erg cm$^{-2}$ s$^{-1}$}
\newcommand{\ecssr}{erg cm$^{-2}$ s$^{-1}$ sr$^{-1}$}

\newcommand{\um}{$\mu$m}                                 
\newcommand{\molh}{H$_{2}$}                              

\newcommand{\water}{H$_{2}$O}

\newcommand{\lsun}{$L_{\odot}$}                          
\newcommand{\msun}{$M_{\odot}$}

\newcommand{\mdot}{{\it \.{M}}}
\newcommand{\msunyr}{$M_{\odot} \, {\rm yr}^{-1}$}

\newcommand{\gapprox}{$\stackrel {>}{_{\sim}}$}   

\newcommand{\about}{$\sim$}                       
\newcommand{\powten}[1]{10$^{#1}$}
\newcommand{\ctwo}{[C\,{\sc ii}]\,157\,$\mu$m}    

\newcommand{\oishort}{[O\,{\sc i}]\,63\,$\mu$m}
\newcommand{\oilong}{[O\,{\sc i}]\,145\,$\mu$m}
\newcommand{\oiiilong}{[O\,{\sc iii}]\,88\,$\mu$m}
\newcommand{\oiiishort}{[O\,{\sc iii}]\,53\,$\mu$m}

\newcommand{\niishort}{[N\,{\sc ii}]\,122\,$\mu$m}

\newcommand{\oknot}{{\rm O}$^0$}

\newcommand{\cplus}{{\rm C}$^+$}

\newcommand{\av}{$A_{\rm V}$}                     

\newcommand{\scc}{${\rm Serpens \,  cloud \,  core}$}

\newcommand{\fir}{FIR}
\newcommand{\pdr}{PDR}
\newcommand{\lte}{LTE}
\newcommand{\lvg}{LVG}
\newcommand{\esa}{ESA}
\newcommand{\nasa}{NASA}
\newcommand{\isas}{ISAS}

\newcommand{\iras}{IRAS}

\newcommand{\iso}{ISO}

\newcommand{\lws}{LWS}

\newcommand{\cam}{CAM}
\newcommand{\isocam}{ISO-CAM}
\newcommand{\cvf}{CVF}
\newcommand{\fov}{FOV}

\newcommand{\fwhm}{FWHM}

\newcommand{\nh}{$n({\rm H})$}                     
\newcommand{\nhtwo}{$n({\rm H}_2)$}

\newcommand{\amin}{$^{\prime}$}                   
\newcommand{\asec}{$^{\prime \prime}$}
\newcommand{\adeg}{$^{\circ}$}

\newcommand{\radot}[4]{\mbox{#1$^{\rm h}$#2$^{\rm m}$#3$\stackrel{\rm s}
{_{\bf\cdot}}$#4}}

\newcommand{\decdot}[4]{\mbox{#1$^{\circ}$ #2$^{\prime}$ #3$\stackrel {\prime
\prime}{_{\bf \cdot}}$#4}}

\newcommand{\ltwo}{$\ell^{\rm II}$}

\begin{document}

   \title{
  The ISO-LWS map of the \scc \thanks{ Based
  on observations with \iso, an \esa\ project with instruments
  funded by \esa\ Member States (especially the PI countries: France, Germany,
  the Netherlands and the United Kingdom) and with the participation of
  \isas\ and \nasa.}
          }

   \subtitle{II. The line spectra}

   \author{Bengt Larsson\inst{1}        \and
           Ren\'e Liseau\inst{1}        \and
           Alexander B. Men'shchikov\inst{1,\,2}
           }

   \offprints{B. Larsson}

   \institute{Stockholm Observatory, SCFAB, Roslagstullsbacken 21, SE-106 91 Stockholm, Sweden \\
              \email{bem@astro.su.se, rene@astro.su.se}
  \and
        Max-Planck-Institut f\"ur Radioastronomie, Auf dem H\"ugel, Bonn, Germany \\
             \email{sasha@mpifr-bonn.mpg.de}
        }

\date{Received date: \hspace{3cm}Accepted date:}


\abstract{We present spectrophotometric \iso\ imaging with the \lws\ and the \cam-\cvf\
of the Serpens molecular cloud core.
The \lws\ map is centred on the far infrared and submillimetre source FIRS\,1/SMM\,1
and its size is 8\amin\,$\times$\,8\amin.
The fine structure line emission in \oishort\ and \ctwo\ is extended on the arcminute scale
and can be successfully modelled to originate in a \pdr\ with  $G_0 = 15 \pm 10$ and
$n$(\molh) in the range of $(10^4 - 10^5)$\,\cmthree. Extended emission might also be
observed in the rotational line emission of \water\ and high-$J$ CO. However, lack of
sufficient angular resolution prevents us from excluding the possibility that the
emssion regions of these lines are point like, which could be linked to the embedded
objects SMM\,9/S\,68 and SMM\,4.
Toward the Class\,0 source SMM\,1, the \lws\ observations reveal, in addition
to fine structure line emission, a rich spectrum of molecular lines, superposed
onto a strong, optically thick dust continuum (Larsson et al. 2000). The sub-thermally
excited and optically thick CO, \water\ and OH lines are tracing an about
\powten{3}\,AU source with temperatures higher than 300\,K and densities above
\powten{6}\,\cmthree\ ($M=0.01$\,\msun). The molecular abundances, $X=N({\rm mol})/N$(\molh), are
$X=(1,\,0.1,\,0.02,\,\ge 0.025) \times 10^{-4}$ for CO, \water, OH and $^{13}$CO, respectively.
Our data are consistent with an ortho-to-para ratio of 3 for \water.
OH appears highly overabundant, which we tentatively ascribe to an enhanced (X-ray) ionisation
rate in the \scc\ ($\zeta \gg 10^{-18}\,{\rm s}^{-1}$). We show that geometry is of concern
for the correct interpretation of the data and
based on 2D-radiative transfer modelling of the disk/torus around SMM\,1, which successfully
reproduces the entire observed SED and the observed line profiles of low-to-mid-$J$
CO isotopomers, we can exclude the disk to be the source of the \lws-molecular line emission.
The same conclusion applies to models of dynamical collapse (`inside-out' infall).
The 6\asec\ pixel resolution of the \cam-\cvf\ permits us to see that the region
of rotational \molh\ emission is offset from SMM\,1 by 30\asec, at position
angle 340\adeg, which is along the known jet flow from the Class\,0 object. This \molh\
gas is extinguished by \av\,=\,4.5\,mag and at a temperature of \powten{3}\,K, which
suggests that the heating of the gas is achieved through relatively slow shocks.
Although we are not able to establish any firm conclusion regarding the detailed
nature of the shock waves, our observations of the molecular line emission from SMM\,1
are to a limited extent explainable in terms of an admixture of J-shocks and of C-shocks,
the latter with speeds of about (15--20)\,\kms, whereas dynamical infall is not directly revealed by our data.
  \keywords{ISM: individual objects: \scc, FIRS\,1/SMM\,1  -- abundances -- molecules -- clouds --
                 jets and outflows -- Stars: formation}
}
  \maketitle
%

\section{Introduction}

The Serpens dark cloud is currently forming a dense cluster
of low to intermediate mass stars (Strom et al. 1976, Kaas 1999). This star forming complex
is situated in the inner Galaxy, not very far from the direction toward the Galactic Centre
(\ltwo\,=\,32\adeg). At the distance of 310\,pc (de\,Lara et al. 1991), the \scc\
is only 30\,pc from the nominal galactic plane, i.e. well within the scale height of the
molecular gas of about 80\,pc (Dame et al. 1987). Not totally unexpected,
\iras-observations revealed intense and patchy far infrared (\fir) emission on a sloping background.
Earlier observations had discovered discrete \fir\ sources of relatively low luminosity
(Nordh et al. 1982, Harvey et al. 1984).

The empirical classification scheme developed by Lada \& Wilking (1984), and later
extended by Andr\'e et al. (1993), is based on the Spectral Energy Distribution (SED)
of the young stellar objects. In a previous paper (Larsson et al. 2000), we were discussing
the SEDs of the submm-sources in the \scc\ and for the dominating source, SMM\,1
(also known as Serpens FIRS\,1; Harvey et al. 1984),
we reached the conclusion that its SED classifies it as Class\,0. The existence of a circumstellar disk
has been announced by Brown et al. (2000). An immediate question is then, whether SMM\,1 shows any
detectable or deducable spectroscopic evidence of disk accretion and/or of dynamical infall.
Over a 0.2\,pc (about 2\amin) region toward the \scc, Williams \& Myers (2000) reported signs of infall.
Observations toward SMM\,1 of  molecular line profiles (Mardones et al. (1997) and Gregersen et al. (1997)),
have not so far been able to reveal a clear `collapse signature'. This is possibly because of confusion with
the known mass outflow activity of the source (Rodr\'{\i}guez et al. 1989, Eiroa et al. 1992,
McMullin et al. 1994, White et al. 1995, Davis et al. 1999, Williams \& Myers 2000, Testi et al. 2000) 
and/or different velocity components in the could complex.
On the other hand, Hogerheijde et al. (1999) found, from continuum interferometric observations,
the density profile of the source to be consistent with the theoretical expectation of a collapsing cloud.

The earlier molecular line results were based on observations of low excitation transitions, which
are not particularly (or not all) sensitive to the conditions expected to prevail in the deeper
layers of the source. In this paper, we present spectral line data of the \scc, both for
a spatial map and for pointed deep integrations, which contain lines also of very high excitation.
These are potentially better suited to `penetrate' to regions which were previously hidden from view.
Our interpretation of the results will especially focus on the evolution of this star forming complex.

In Sect.\,2, we reiterate the LWS observations and a summary is given for the data
reductions, whereas a more detailed account is provided in Appendix\,A. The resulting
line spectra are presented in Sect.\,3. These results are discussed at some depth in
Sect.\,4. We first exploit relatively simple analytical methods, which result in spatially
averaged properties. These should be useful to limit the parameter space for
more sophisticated numerical modelling. This is done for the transfer of both continuum
and line radiation, and these models lead to some valuable conclusions. In a summarising discussion,
we make an attempt to bring these various pieces of information into a coherent physical picture.
Finally in Sect.\,5, we summarise our main conclusions from this work.

\section{Observations and data reductions}

\subsection{The \iso\ observations}

A $5 \times 5$ spectrophotometric map in the far-infrared (\fir) of the \scc\ was obtained with the
Long-Wavelength Spectrometer (\lws; 43 -- 197\,\um, $R_{\lambda}=140 - 330$)
on board the Infrared Space Observatory (\iso) on October 21, 1996.
The \iso-project is described by Kessler et al. (1996) and
the \lws\ is described by Clegg et al. (1996) and Swinyard et al. (1996).

The formal map centre, viz. $\alpha$ = \radot{18}{29}{50}{29} and $\delta$ = \decdot{1}{15}{18}{6},
epoch J\,2000, coincides to within 10\asec\ with the position of the
sub-millimetre source SMM\,1. The pointing accuracy in the map is determined as 1\asec\ (rms).
The spacings between positions in the map, oriented along the equatorial coordinates, are
100\asec. The size of the \lws-map is 8\amin\,$\times$\,8\amin, corresponding to
$(0.7\times 0.7 = 0.5)$\,$D_{310}^2$\,\pcq, where $D_{310}$ denotes the distance
to the \scc\ in units of 310\,pc.

At each map-point the grating of the \lws\ was scanned 6 times in fast mode, oversampling
the spectral resolution at twice the Nyquist rate. Each position was observed for nearly 15\,min.
The centre position was re-observed half a year later on April 15, 1997, for a considerably
longer integration time (24 spectral scans).

In addition, imaging spectrophotometric data obtained with the Continuous Variable Filter
(\cvf; 5 -- 16.5\,\um, $32 \times 32$\,pxl, pixel-\fov\,=\,6\asec, $R_{\lambda}$\,\gapprox\,35) of \isocam\
(Cesarsky et al. 1996) were also analysed. The detailed observing log is given in our
previous paper (Larsson et al. 2000).

\subsection{The reduction of the \iso\ data}

\subsection{\lws\ data}

The \lws\ data reductions were done with the interactive analysis package LIA. This
pipeline processing was done in two ways. For the long wave
detectors LW\,1 -- LW\,5, the Relative Spectral Response Functions (RSRFs) of the most recent version
(OLP\,10) provided an overall better agreement than the older OLP\,8 for the detector inter-calibration.
However, the OLP\,10 RSRFs also contained a number of strong features, which resulted in spurious `lines'
in the reduced spectra (Appendix\,A). We decided therefore to use the RSRFs of OLP\,8,
which did not show this behaviour,
but scaled to the OLP\,10 absolute levels. For the short-wave detectors SW\,1 -- SW\,5, no such difference
between OLP\,8 and OLP\,10 was apparent and consequently we used the latest version. A detailed
account for the first steps in the \lws\ reduction is given in Appendix\,A.

Subsequent post-pipeline processing used the package ISAP.
At each position, the individual \lws\ scans for each of the 10 detectors were
examined, `deglitched' and averaged. Corrections were applied to the
`fringed' spectra in the map. The fringing indicates that the emission is
extended and/or that point sources were not on the optical axis of the \lws, i.e.
the radial distance from the optical axis was typically larger than about 25\asec.

The absolute flux calibration is better than about 30\% (Swinyard et al. 1996).
Overlapping spectral regions of adjacent detectors were generally within 10\%
(`detector stitching' uncertainty). At 60\,\um\ and 100\,\um, the correspondance with
broad-band \iras\ data is even better than 10\% (Larsson et al. 2000).

\subsection{\isocam-\cvf\ data}

\begin{figure}[t]
  \resizebox{\hsize}{!}{
  \rotatebox{00}{\includegraphics{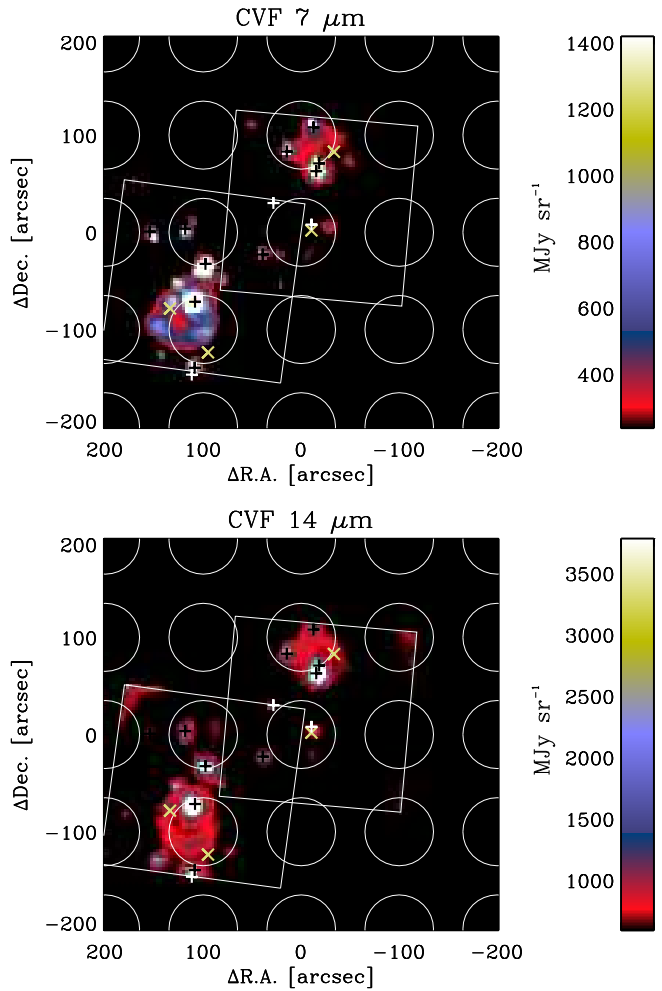}}
                        }
  \caption{The from \cvf\ frames synthesised images at 7\,\um\ and 14\,\um. From comparison with
  \cam\ broad-band observations at these wavelengths and deep K-images the point sources were identified
  (see: Kaas 1999). These are shown by plus-signs (white and black) and were used to align the \cvf\
  frames with the equatorial coordinate system in the sky. Major SMM sources in the field are shown
  by yellow crosses. The \lws\ map pointings are indicated by the contours of the \fwhm\ of the beam
  (white circles).}
  \label{cvf_broadband}
\end{figure}

The \isocam\ data were reduced with the CIA programs using OLP\,8.4. This included dark current
correction, transient correction and deglitching and flat field corrections. The calibrations
were based on OLP\,5.4. The proper alignment of the \cvf\ fields was achieved with the aid of
point sources identified in both \cvf\ and broad-band (7\,\um\ and 14\,\um) \cam\ observations
(Fig.\,\ref{cvf_broadband}).

\section{Results}

\subsection{Spectral imaging in \molh}

\begin{figure*}
  \resizebox{\hsize}{!}{
  \rotatebox{90}{\includegraphics{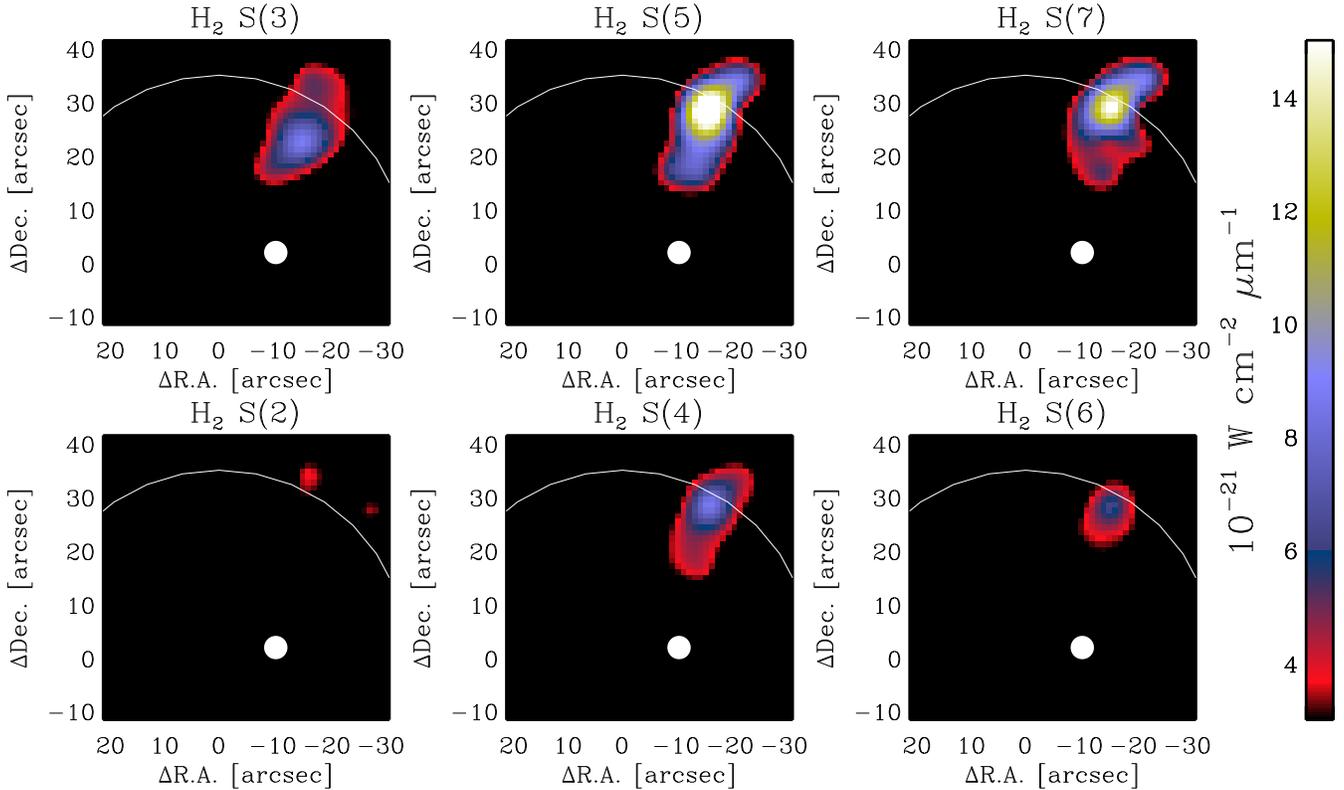}}
                        }
  \caption{\cvf\ images toward SMM\,1 in 3 ortho-\molh\ (upper frames) and 3 para-\molh\ (lower row) lines.
  Offsets are in arcsec and relative to the centre position (0, 0) of the \lws\ map. The white arcs outline
  the \fwhm\ contour of the \lws\ beam (70\asec) and the white spot designates the position of SMM\,1. The
  actual \cvf\ pixels are square 6\asec.}
  \label{h2_map}
\end{figure*}

Lines from purely rotationally excited \molh\ are clearly discernible in the \cvf-image.
In particular, all lines which fall in the observed spectral range, were detected and,
in Table\,\ref{tab_h2}, their observed fluxes are listed. These refer to
the 3 para-\molh\ lines 0--0\,S(2)\,12.3, S(4)\,8.0 and S(6)\,6.1\,\um\ and the 3
ortho-\molh\ lines 0--0\,S(3)\,9.7, S(5)\,6.9 and S(7)\,5.5\,\um.
Fig.\,\ref{h2_map} shows maps near SMM\,1 in these six lines, delineating the spatial distribution
of the rotationally excited \molh\ gas along the known jet-flow from SMM\,1
(Eiroa \& Casali 1989, Hodapp 1999). In addition, somewhat weaker \molh\ emission is also observed
near SMM\,3, the discussion of which will be postponed to a future paper.

\begin{table}
\begin{flushleft}
 \caption{\label{tab_h2} Observed H$_2$ fluxes near SMM\,1}
 \begin{tabular}{lll}
  \hline
   Transition       & Peak$^{\dagger}$           & Total$^{\ddagger}$      \\
                    & ($10^{-21}$ W cm$^{-2}$)   & ($10^{-20}$ W cm$^{-2}$) \\
  \noalign{\smallskip}
  \hline
  \noalign{\smallskip}
  0-0 S(2)          &  \phantom{1}$2.9  \pm  0.9$ & $1.5  \pm  0.2$ \\
  0-0 S(3)          &  \phantom{1}$5.2  \pm  0.6$ & $4.8  \pm  0.4$ \\
  0-0 S(4)          &  \phantom{1}$8.8  \pm  1.0$ & $3.6  \pm  0.3$ \\
  0-0 S(5)          &             $19.8 \pm  1.2$ & $7.0  \pm  0.5$ \\
  0-0 S(6)          &  \phantom{1}$6.2  \pm  0.8$ & $0.7  \pm  0.2$ \\
  0-0 S(7)          &             $15.6 \pm  2.8$ & $5.8  \pm  0.6$ \\
  1-0 S(1)$^{\amalg}$ & \phantom{1}2.5            &  1.0            \\
  \noalign{\smallskip}
  \hline
  \end{tabular}
\end{flushleft}
Notes to the table: \\
$^{\dagger}$ Peak emission into a 36 arcsec$^2$ pixel. \\
$^{\ddagger}$ Total emission over a source of 350 arcsec$^2$. \\
$^{\amalg}$ Estimated from Fig.\,3 of Eiroa and Casali (1989). \\
\end{table}

\subsection{\lws-maps of \oishort, \ctwo, ortho-\water\,179.5\,\um\ and CO\,186\um}

\begin{figure*}
  \resizebox{\hsize}{!}{
  \rotatebox{90}{\includegraphics{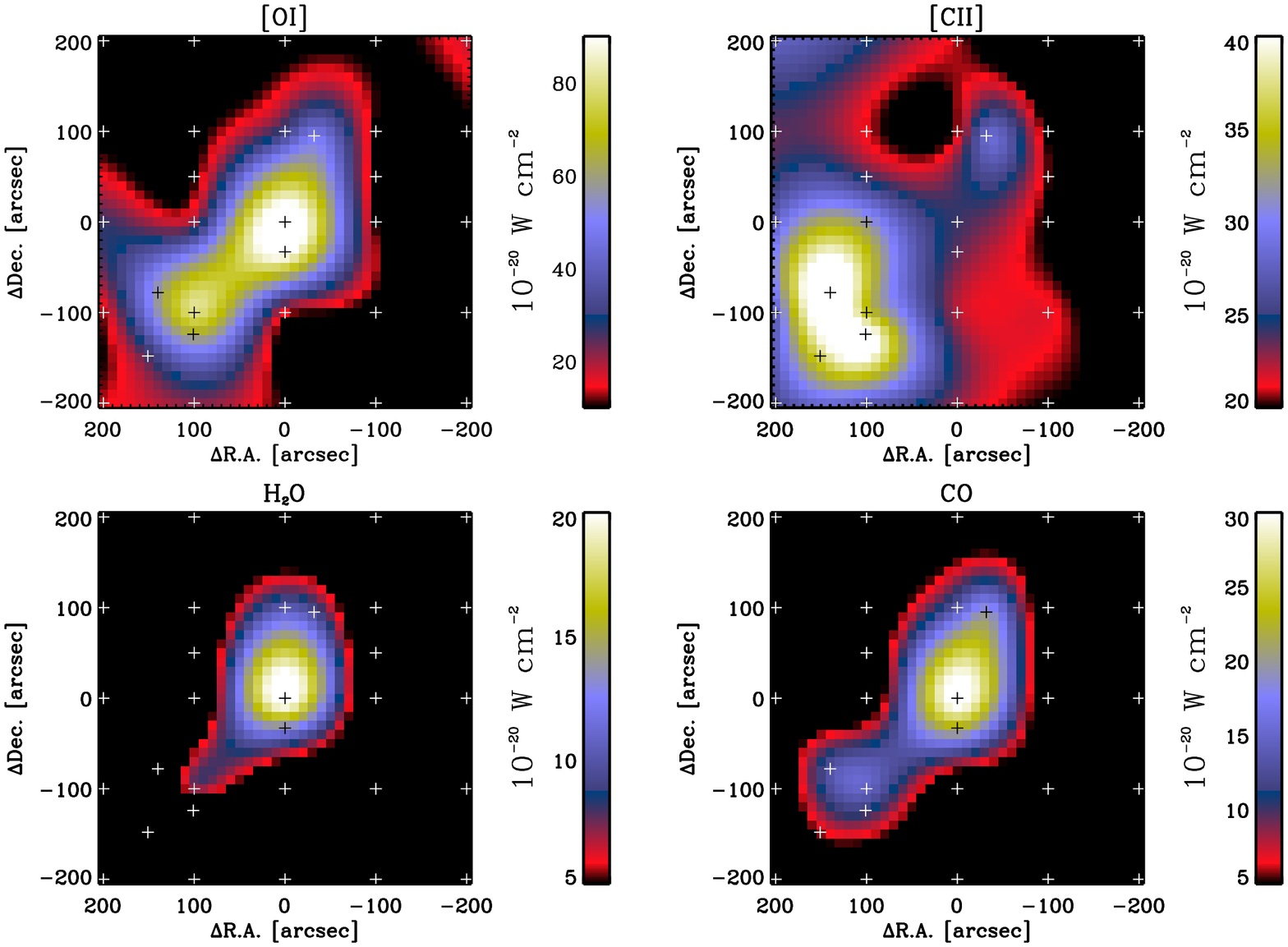}}
                        }
  \caption{Maps of integrated line fluxes of \oishort, \ctwo, \water\,($2_{12}-1_{01}$) and
  CO\,($14-13$), with individual flux levels indicated by the bars next to each frame. The crosses
  refer to the positions toward which the \lws\ spectra have been obtained.}
  \label{mol_map}
\end{figure*}

Within the $8^{\prime}\times 8^{\prime}$ map observed with the \lws, the spatial distribution
of the fine structure lines \oishort\ and \ctwo\ is shown in Fig.\,\ref{mol_map} together with that of
the rotational lines of \water\,($2_{12}-1_{01}$) (connecting to the ground state) and CO\,($J=14-13$)
(lowest $J-$transition admitted and close in wavelength to the \water\ line).

Like the emission in \ctwo, \oishort\ emission is observed in each position of our map, but with
maximum intensity at the position of SMM\,1. Secondary maxima
are also seen along a ridge toward SMM\,4/3/8 in the southeast-to-east. In addition, the emission is
extended toward SMM\,9/S\,68\,NW. The overall \oishort\ distribution is strikingly similar to that seen
in the mm-regime in a variety of high density tracing molecules (McMullin et al. 2000).
Very interesting is also the fact that, just in the northwest corner of the map, the shocked flow
HH\,460 (Davis et al. 1999, Ziener \& Eisl\"offel 1999) seems to be discernible.

In \ctwo, maximum emission is seen toward S\,68 (Sharpless 1959) and the cluster in the southeast, whereas
SMM\,1 is inconspicuous in this map. To some extent, the \ctwo-map is an inverted image of that
seen in \oishort.

The molecular emision in the 179.5\,\um\ line of ortho-\water\ and in the 186\,\um\ line of CO
is strongest toward SMM\,1, but displays also some extensions along the \oishort\ ridge, i.e.
toward the northwest and the southeast. The spatial resolution is not sufficient, however,
to decide whether this emission is extended or due to the numerous SMM-sources in the region
(cf. Fig.\,\ref{mol_map}).

\subsection{The line spectrum of SMM\,1 from 43 to 197\,\um}

\begin{figure*}
  \resizebox{\hsize}{!}{
  \rotatebox{90}{\includegraphics{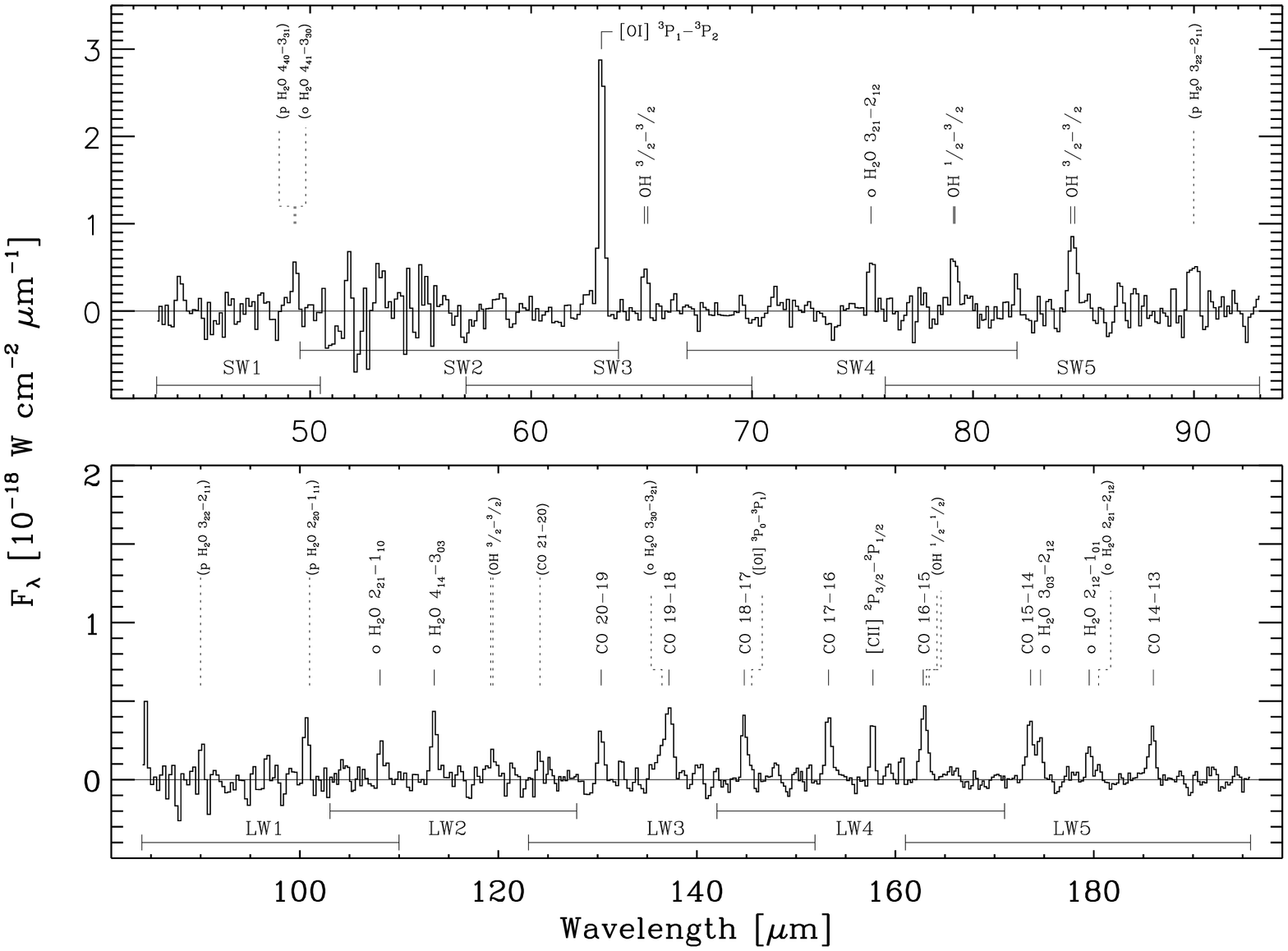}}
                        }
  \caption{The \lws\ spectrum (43 to 197\,\um) of Serpens SMM\,1 with line positions indicated. For
  clarity the strong continuum (Larsson et al. 2000) has been subtracted. The
  originally four times oversampled data have been rebinned to 0.1\,\um\ for the short
  wavelength (SW) and to 0.2\,\um\ for the long wavelength (LW) spectral range. Also shown,
  below the spectrum, is the extent of the individual \lws\ detectors.}
  \label{smm1_spec}
\end{figure*}

The continuum subtracted line spectrum of SMM\,1 from 43 to 197\,\um\ is displayed in
Fig.\,\ref{smm1_spec}, revealing
a plethora of emission lines from both atomic and molecular species. In the figure, also the
wavelength coverage of the ten individual LWS detectors is indicated (cf. Sect.\,2.2).
The identification of the lines and their parameters, as obtained from profile fitting,
are listed in Tables\,\ref{tab_atom} to \ref{tab_OH}. Listed are the identified species, the upper and lower states
and the rest wavelength of the transition. This is followed by the observed wavelength with the
measurement error and the difference between the observed and the rest wavelength. Then the line width and
the line flux with fitting-errors, respectively, are tabulated and the error estimate for the line flux,
i.e. $\Delta F = \sqrt{\sigma_{\rm l}^2 + \sigma_{\rm c}^2}$ where $\sigma_{\rm l}$ is the fitting error and
$\sigma_{\rm c}$ is the rms-level of the surrounding continuum integrated over one spectral resolution element.
In the last two columns, the type of the Gauss fitting (single or multiple component) and the LWS detector are given.

Out of a total of 8 measured CO lines, clear detections are found for
7 transitions.
In addition, we identfy 5 (possibly 8) lines of ortho-\water, and 1 (possibly 2) of
para-\water. Further, 4 lines of OH are clearly present in the spectrum,
whereas for two more lines this status is less clear. Aside from the line emission from molecules,
lines from \oknot\ and \cplus\ are also present. Possible implications from these results
will be discussed in the following sections.

\begin{table*}
 \caption{\label{tab_atom} Line fluxes of atoms and ions toward SMM\,1}
  \begin{tabular}{lc ccc c c c cl}
  \hline
  \noalign{\smallskip}
   Ion id      & Transition & \multicolumn{4}{c}{Wavelength}                       & \multicolumn{2}{c}{Flux$\times 10^{19}$}   & Single/ & \lws \\
               &            & \multicolumn{4}{c}{($\mu$m)}                         & \multicolumn{2}{c}{(W cm$^{-2}$)}          & Multi   & Detector \\
               &            &$\lambda$& $\lambda_{\rm obs}$& $\Delta \lambda$ & $\sigma_{\lambda}$    &  $F$     & $\Delta F$   & fit     &  \\
  \noalign{\smallskip}
  \hline
  \noalign{\smallskip}
   $[$CII$]$  & $^2$P$_{3/2}$ - $^2$P$_{1/2}$ & 157.74 & 157.72 $\pm$ 0.02 & $-$0.02 & 0.46 $\pm$ 0.03 & 2.07 $\pm$ 0.21 & 0.35 & S & LW 4 \\
              &                               &        & 157.72 $\pm$ 0.04 & $-$0.02 & 0.60 Fix        & 2.31 $\pm$ 0.20 & 0.34 & S & LW 4 \\
   $[$OI$]$   & $^3$P$_0$ - $^3$P$_1$         & 145.53 & 145.53 Fix        & +0.00   & 0.60 Fix        & 0.37 $\pm$ 0.84 & 0.90 & M & LW 4 \\
              &                               &        & 145.53 Fix        & +0.00   & 0.60 Fix        & 0.76 $\pm$ 0.83 & 0.89 & M & LW 3 \\
              & $^3$P$_1$ - $^3$P$_2$         &  63.18 &  63.19 $\pm$ 0.01 & +0.01   & 0.27 $\pm$ 0.01 &  9.5 $\pm$ 0.7  & 0.74 & S & SW 3 \\
              &                               &        &  63.19 $\pm$ 0.01 & +0.01   & 0.29 Fix        &  9.8 $\pm$ 0.4  & 0.46 & S & SW 3 \\
              &                               &        &  63.17 $\pm$ 0.01 & $-$0.01 & 0.28 $\pm$ 0.01 & 10.3 $\pm$ 0.9  & 0.96 & S & SW 2 \\
              &                               &        &  63.17 $\pm$ 0.01 & $-$0.01 & 0.29 Fix        & 10.5 $\pm$ 0.5  & 0.60 & S & SW 2 \\
   $[$NII$]$  & $^3$P$_2$ - $^3$P$_1$         & 121.90 &                   &         &                 &                 & 0.29 & S & LW 2 \\
   $[$OIII$]$ & $^3$P$_1$ - $^3$P$_0$         &  88.36 &                   &         &                 &                 & 0.47 & S & LW 1 \\
              &                               &        &                   &         &                 &                 & 0.32 & S & SW 5 \\
              & $^3$P$_2$ - $^3$P$_1$         &  51.82 &                   &         &                 &                 & 0.73 & S & SW 2 \\

  \noalign{\smallskip}
  \hline
  \noalign{\smallskip}
  \end{tabular}
\end{table*}

\begin{table*}
 \caption{\label{tab_co} CO line measurements in the spectrum of SMM\,1}
  \begin{tabular}{l ccc c c c ll}
  \hline
  \noalign{\smallskip}
Transition& \multicolumn{3}{c}{Wavelength} & Width & Flux$\times 10^{19}$ & Err$\times 10^{19}$ & Single/ & \lws\\
         & \multicolumn{3}{c}{($\mu$m)} &  ($\mu$m)& (W cm$^{-2}$)        & (W cm$^{-2}$)            & Multi   & Detector \\
{\small ($J \rightarrow J-1$)} & $\lambda$ & $\lambda_{\rm obs}$ & $\Delta \lambda$ & $\sigma_{\lambda}$ & $F$ & $\Delta F$ & fit & \\
  \noalign{\smallskip}
  \hline
  \noalign{\smallskip}
 $14-13$ & 186.00 & 185.88 $\pm$ 0.03 & $-$0.12 & 0.76 $\pm$ 0.03 & {\underline {2.64 $\pm$ 0.30}} & 0.37 & S & LW 5 \\
         &        & 185.90 $\pm$ 0.03 & $-$0.10 & 0.60 Fix        &              2.32 $\pm$ 0.16   & 0.27 & S & LW 5 \\
 $15-14$ & 173.63 & 173.60 $\pm$ 0.03 & $-$0.03 & 0.81 $\pm$ 0.03 &              3.34 $\pm$ 0.28   & 0.35 & S & LW 5 \\
         &        & 173.59 $\pm$ 0.03 & $-$0.04 & 0.60 Fix        &              2.81 $\pm$ 0.21   & 0.30 & S & LW 5 \\
         &        & 173.58 $\pm$ 0.03 & $-$0.05 & 0.60 Fix        & {\underline {2.77 $\pm$ 0.50}} & 0.55 & M & LW 5 \\
 $16-15$ & 162.81 & 162.90 $\pm$ 0.02 & +0.09   & 0.98 $\pm$ 0.02 &              4.58 $\pm$ 0.19   & 0.28 & S & LW 5 \\
         &        & 162.92 $\pm$ 0.04 & +0.12   & 0.60 Fix        &              3.38 $\pm$ 0.34   & 0.40 & S & LW 5 \\
         &        & 162.81 Fix        &         & 0.60 Fix        & {\underline {2.67 $\pm$ 0.50}} & 0.54 & M & LW 5 \\
         &        & 162.93 $\pm$ 0.02 & +0.12   & 0.59 $\pm$ 0.02 &              3.19 $\pm$ 0.18   & 0.36 & S & LW 4 \\
         &        & 162.93 $\pm$ 0.04 & +0.12   & 0.60 Fix        &              3.21 $\pm$ 0.10   & 0.33 & S & LW 4 \\
         &        & 162.81 Fix        & +0.00   & 0.60 Fix        & {\underline {2.86 $\pm$ 0.34}} & 0.46 & M & LW 4 \\
 $17-16$ & 153.27 & 153.28 $\pm$ 0.02 & +0.01   & 0.64 $\pm$ 0.02 & {\underline {2.81 $\pm$ 0.20}} & 0.34 & S & LW 4 \\
         &        & 153.27 $\pm$ 0.02 & +0.00   & 0.60 Fix        &              2.71 $\pm$ 0.11   & 0.30 & S & LW 4 \\
 $18-17$ & 144.78 & 144.73 $\pm$ 0.02 & $-$0.05 & 0.57 $\pm$ 0.03 &              2.55 $\pm$ 0.31   & 0.44 & S & LW 4 \\
         &        & 144.75 $\pm$ 0.03 & $-$0.03 & 0.60 Fix        &              2.63 $\pm$ 0.16   & 0.36 & S & LW 4 \\
         &        & 144.78 Fix        &         & 0.60 Fix        & {\underline {2.50 $\pm$ 0.22}} & 0.39 & M & LW 4 \\
         &        & 144.78 $\pm$ 0.02 & +0.00   & 0.50 $\pm$ 0.02 &              2.67 $\pm$ 0.28   & 0.42 & S & LW 3 \\
         &        & 144.79 $\pm$ 0.03 & +0.01   & 0.60 Fix        & {\underline {2.94 $\pm$ 0.23}} & 0.38 & S & LW 3 \\
         &        & 144.78 Fix        &         & 0.60 Fix        &              2.94 $\pm$ 0.32   & 0.44 & M & LW 3 \\
 $19-18$ & 137.20 & 137.19 $\pm$ 0.03 & $-$0.01 & 0.80 $\pm$ 0.04 &              3.39 $\pm$ 0.41   & 0.53 & S & LW 3 \\
         &        & 137.21 $\pm$ 0.04 & +0.01   & 0.60 Fix        &              2.85 $\pm$ 0.28   & 0.43 & S & LW 3 \\
         &        & 137.20 Fix        &         & 0.60 Fix        & {\underline {2.83 $\pm$ 0.30}} & 0.45 & M & LW 3 \\
 $20-19$ & 130.37 & 130.35 $\pm$ 0.03 & $-$0.02 & 0.68 $\pm$ 0.03 &              2.48 $\pm$ 0.32   & 0.48 & S & LW 3 \\
         &        & 130.34 $\pm$ 0.03 & $-$0.03 & 0.60 Fix        & {\underline {2.33 $\pm$ 0.18}} & 0.39 & S & LW 3 \\
 $21-20$ & 124.19 & 124.21 $\pm$ 0.05 & +0.02   & 0.48 $\pm$ 0.06 &              0.98 $\pm$ 0.32   & 0.48 & S & LW 3 \\
         &        & 124.20 $\pm$ 0.08 & +0.01   & 0.60 Fix        & {\underline {1.08 $\pm$ 0.23}} & 0.42 & S & LW 3 \\
         &        & 124.10 $\pm$ 0.05 & $-$0.09 & 0.52 $\pm$ 0.06 &              1.17 $\pm$ 0.31   & 0.42 & S & LW 2 \\
         &        & 124.10 $\pm$ 0.06 & $-$0.09 & 0.60 Fix        & {\underline {1.25 $\pm$ 0.18}} & 0.34 & S & LW 2 \\
 $22-21$ & 118.58 &                   &         &                 &                                & 0.29 & S & LW 2 \\
  \noalign{\smallskip}
  \hline
  \noalign{\smallskip}
  \end{tabular}
  \\
Notes to the table: \\
$\Delta F = \sqrt{\sigma_{\rm l}^2 + \sigma_{\rm c}^2}$ where $\sigma_{\rm l}$ is the fitting error and
$\sigma_{\rm c}$ is the rms-level of the surrounding continuum integrated over one spectral resolution element,
i.e. 0.29 $\mu$m and 0.60 $\mu$m for the SW and LW detectors, respectively.
S = Single line component fit; M = multi component line fit.
Underlined flux values were used in the rotation diagram analysis and are shown in Fig.\,\ref{co_rot}.
\end{table*}

\begin{table*}
 \caption{\label{tab_H2O} \water\ line measurements in the spectrum of SMM\,1}
  \begin{tabular}{l ccc c c c ll}
  \hline
  \noalign{\smallskip}
  Transition & \multicolumn{3}{c}{Wavelength} & Width  & Flux$\times 10^{19}$  & Err$\times 10^{19}$  & Single/ & \lws  \\
         & \multicolumn{3}{c}{($\mu$m)} & ($\mu$m) & (W cm$^{-2}$) & (W cm$^{-2}$)        & Multi   & Detector \\
         & $\lambda$ & $\lambda_{\rm obs}$ & $\Delta \lambda$ & $\sigma_{\lambda}$& $F$ & $\Delta F$     & fit     &  \\
  \noalign{\smallskip}
  \hline
  \noalign{\smallskip}
   {\underline{ortho-H$_2$O}} & &  & & & & & & \\
  \noalign{\smallskip}
   $2_{21}-2_{12}$  & 180.49 & 180.49 $\pm$ 0.03 & +0.00 & 0.45 $\pm$ 0.03 & 0.53 $\pm$ 0.11 & 0.24 & S & LW 5 \\
                    &        & 180.49 $\pm$ 0.05 & +0.00 & 0.60 Fix        & 0.60 $\pm$ 0.01 & 0.23 & S & LW 5 \\
   $2_{12}-1_{01}$  & 179.53 & 179.55 $\pm$ 0.04 & +0.02 & 0.57 $\pm$ 0.04 & 1.40 $\pm$ 0.24 & 0.32 & S & LW 5 \\
                    &        & 179.55 $\pm$ 0.04 & +0.02 & 0.60 Fix        & 1.43 $\pm$ 0.13 & 0.25 & S & LW 5 \\
   $3_{03}-2_{12}$  & 174.63 & 174.63 $\pm$ 0.03 & +0.00 & 0.70 $\pm$ 0.03 & 2.15 $\pm$ 0.24 & 0.33 & S & LW 5 \\
                    &        & 174.63 $\pm$ 0.03 & +0.00 & 0.60 Fix        & 1.97 $\pm$ 0.14 & 0.26 & S & LW 5 \\
                    &        & 174.63 $\pm$ 0.06 & +0.00 & 0.60 Fix        & 1.93 $\pm$ 0.66 & 0.71 & M & LW 5 \\
   $3_{30}-3_{21}$  & 136.49 & 136.52 $\pm$ 0.09 & +0.03 & 0.64 $\pm$ 0.11 & 0.85 $\pm$ 0.34 & 0.48 & S & LW 3 \\
                    &        & 136.51 $\pm$ 0.06 & +0.02 & 0.60 Fix        & 0.81 $\pm$ 0.11 & 0.35 & S & LW 3 \\
                    &        & 136.51 Fix        & +0.02 & 0.60 Fix        & 0.77 $\pm$ 0.82 & 0.88 & M & LW 3 \\
   $4_{14}-3_{03}$  & 113.54 & 113.54 $\pm$ 0.04 & +0.00 & 0.61 $\pm$ 0.04 & 2.96 $\pm$ 0.39 & 0.52 & S & LW 2 \\
                    &        & 113.54 $\pm$ 0.03 & +0.00 & 0.60 Fix        & 2.93 $\pm$ 0.21 & 0.41 & S & LW 2 \\
   $2_{21}-1_{10}$  & 108.07 & 108.13 $\pm$ 0.05 & +0.06 & 0.39 $\pm$ 0.02 & 1.29 $\pm$ 0.40 & 0.50 & S & LW 2 \\
                    &        & 108.18 $\pm$ 0.07 & +0.11 & 0.60 Fix        & 1.58 $\pm$ 0.31 & 0.45 & S & LW 2 \\
                    &        & 107.83 $\pm$ 0.03 & $-$0.25 & 0.51 $\pm$ 0.05 & 1.60 $\pm$ 0.24 & 0.52 & S & LW 1 \\
                    &        & 107.84 $\pm$ 0.05 & $-$0.24 & 0.60 Fix        & 1.79 $\pm$ 0.22 & 0.51 & S & LW 1 \\
   $3_{21}-2_{12}$  &  75.38 &  75.40 $\pm$ 0.03 & +0.02 & 0.29 $\pm$ 0.03 & 2.05 $\pm$ 0.51 & 0.60 & S & SW 4 \\
                    &        &  75.40 $\pm$ 0.03 & +0.02 & 0.29 Fix        & 2.03 $\pm$ 0.28 & 0.42 & S & SW 4 \\
   $4_{41}-3_{30}$  &  49.34 &  49.32 $\pm$ 0.02 & $-$0.02 & 0.25 $\pm$ 0.02 & 1.69 $\pm$ 0.29 & 0.37 & S & SW 1 \\
                    &        &  49.32 $\pm$ 0.02 & $-$0.02 & 0.29 Fix        & 1.81 $\pm$ 0.18 & 0.29 & M & SW 1 \\
   \noalign{\smallskip}
    {\underline{para-H$_2$O}} & &  & & & & & & \\
   \noalign{\smallskip}
   $2_{20}-1_{11}$  & 100.98 & 100.66 $\pm$ 0.05 & $-$0.32 & 0.57 $\pm$ 0.04 & 2.53 $\pm$ 0.68 & 0.82 & S & LW 1 \\
                    &        & 100.66 $\pm$ 0.06 & $-$0.32 & 0.60 Fix        & 2.60 $\pm$ 0.37 & 0.58 & S & LW 1 \\
   $3_{22}-2_{11}$  &  89.99 &  89.99 $\pm$ 0.05 & +0.00 & 0.44 $\pm$ 0.04 & 2.71 $\pm$ 0.76 & 0.83 & S & SW 4 \\
                    &        &  90.01 $\pm$ 0.04 & +0.02 & 0.29 Fix        & 2.02 $\pm$ 0.42 & 0.53 & S & SW 4 \\
 \noalign{\smallskip}
  \hline
  \noalign{\smallskip}
  \end{tabular}
\end{table*}

\begin{table*}
 \caption{\label{tab_OH} OH line measurements in the spectrum of SMM\,1}
  \begin{tabular}{cc ccc c c c cl}
  \hline
  \noalign{\smallskip}
\multicolumn{2}{c}{$^2\Pi$\,Transition}& \multicolumn{4}{c}{Wavelength} & Flux$\times 10^{19}$ & Err$\times 10^{19}$   & Single/ & \lws    \\
             &                   & \multicolumn{4}{c}{($\mu$m)}   & (W cm$^{-2}$)        & (W cm$^{-2}$)               & Multi   & Detector \\
$\Omega - \Omega^{\prime}$ & $J - J^{\prime}$ & $\lambda$ & $\lambda_{\rm obs}$& $\Delta \lambda$ & $\sigma_{\lambda}$ & $F$ & $\Delta F$& fit & \\
  \noalign{\smallskip}
  \hline
  \noalign{\smallskip}
   $^1\!/_2 - ^1\!\!/_2$ & $^3\!/_2 - ^1\!\!/_2$ & 163.26 & 163.26 Fix        & +0.00 & 0.60 Fix        & 1.39 $\pm$ 0.90 & 0.93 & S & LW 5 \\
                         &                       &        & 163.26 Fix        & +0.00 & 0.60 Fix        & 1.00 $\pm$ 0.85 & 0.91 & S & LW 4 \\
   $^3\!/_2 - ^3\!\!/_2$ & $^5\!/_2 - ^3\!\!/_2$ & 119.34 & 119.38 $\pm$ 0.20 & +0.04 & 1.31 $\pm$ 0.11 & 2.07 $\pm$ 0.49 & 0.56 & S & LW 2 \\
                         &                       &        & 119.42 $\pm$ 0.07 & +0.08 & 0.60 Fix        & 1.28 $\pm$ 0.22 & 0.36 & S & LW 2 \\
   $^3\!/_2 - ^3\!\!/_2$ & $^7\!/_2 - ^5\!\!/_2$ &  84.51 &  84.43 $\pm$ 0.03 & $-$0.08 & 0.39 $\pm$ 0.03 & 2.94 $\pm$ 0.64 & 0.79 & S & LW 1 \\
                         &                       &        &  84.44 $\pm$ 0.08 & $-$0.07 & 0.60 Fix        & 3.49 $\pm$ 0.73 & 0.87 & S & LW 1 \\
                         &                       &        &  84.51 $\pm$ 0.02 & +0.00 & 0.41 $\pm$ 0.02 & 4.02 $\pm$ 0.50 & 0.58 & S & SW 5 \\
                         &                       &        &  84.50 $\pm$ 0.02 & $-$0.01 & 0.29 Fix        & 3.26 $\pm$ 0.31 & 0.42 & S & SW 5 \\
   $^1\!/_2 - ^3\!\!/_2$ & $^1\!/_2 - ^3\!\!/_2$ &  79.15 &  79.14 $\pm$ 0.02 & $-$0.01 & 0.29 $\pm$ 0.02 & 2.84 $\pm$ 0.38 & 0.54 & S & SW 5 \\
                         &                       &        &  79.14 $\pm$ 0.02 & $-$0.01 & 0.29 Fix        & 2.86 $\pm$ 0.21 & 0.43 & S & SW 5 \\
                         &                       &        &  79.17 $\pm$ 0.07 & +0.02 & 0.56 $\pm$ 0.07 & 2.22 $\pm$ 0.80 & 0.89 & S & SW 4 \\
                         &                       &        &  79.14 $\pm$ 0.04 & $-$0.01 & 0.29 Fix        & 1.47 $\pm$ 0.34 & 0.51 & S & SW 4 \\
   $^3\!/_2 - ^3\!\!/_2$ & $^9\!/_2 - ^7\!\!/_2$ &  65.21 &  65.18 $\pm$ 0.03 & $-$0.03 & 0.24 $\pm$ 0.03 & 1.51 $\pm$ 0.45 & 0.47 & S & SW 3 \\
                         &                       &        &  65.18 $\pm$ 0.03 & $-$0.03 & 0.29 Fix        & 1.64 $\pm$ 0.26 & 0.30 & S & SW 3 \\
   $^3\!/_2 - ^3\!\!/_2$ &$^{11}\!/_2 - ^9\!\!/_2$& 53.30 &                   &       &                 &                 & 0.73 & S & SW 2 \\
  \noalign{\smallskip}
  \hline
  \noalign{\smallskip}
  \end{tabular}
  \\
Note to the table: $\lambda$ is an average wavelength for the various fine structure lines.
\end{table*}

\section{Discussion}

\subsection{The atomic line spectrum: CNO}

\subsubsection{\oiiishort, \oiiilong\ and \niishort}

As is evident from Table\,\ref{tab_atom}, only upper limits were obtained for the high ionisation lines
\oiiishort, \oiiilong\ and \niishort. This is consistent with the luminosities derived by
Larsson et al. (2000), indicating the presence of stellar sources generating at best only gentle UV-fields.

We can also exclude the presence of {\it extended} strongly shocked regions. For instance,
if associated with the fast moving objects of Rodr\'{\i}guez et al. (1989),
our data imply that the physical scales of these shocks would be small,
$\ll 1$\asec\ (e.g. for \vs\,\about\,200\,\kms\ and
\no\,\gapprox\,\powten{5}\,\cmthree; see: Cameron \& Liseau 1990; Liseau et al. 1996a).

We can conclude that the degree of ionisation of the atomic gas is generally low.
Lines of low-ionisation species will be discussed in the next sections.

\subsubsection{Extended emission: \ctwo\ and \oishort}

The spatial distribution of the \ctwo\ emission is shown in Fig.\,\ref{mol_map}
from which it is evident that the emission varies within a factor of about two.
The S\,68 nebulosity is pronounced in the \ctwo\ line, making it likely
that its origin is from a photondominated region (\pdr), close to the cloud surface.

\begin{figure}
  \resizebox{\hsize}{!}{
  \rotatebox{90}{\includegraphics{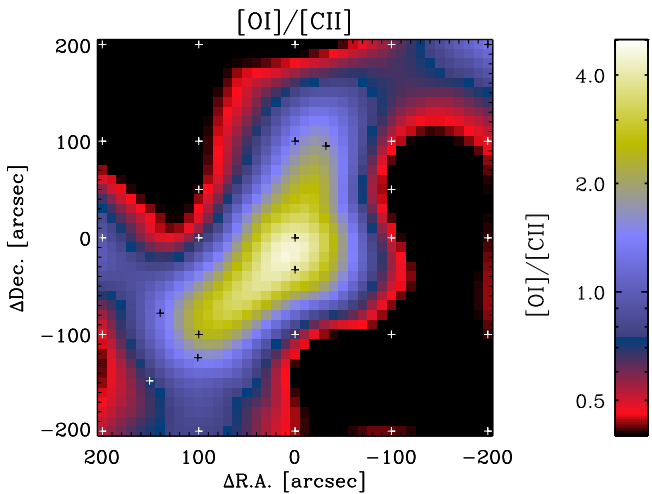}}
                        }
  \caption{Emission line intensity ratio map for \oishort/\ctwo.
  Observed positions are identified by crosses. The observed flux distributions
  of the individual lines are shown in Fig.\,\ref{mol_map}.}
  \label{oicii_ratio}
\end{figure}

This idea can be tested quantitatively by invoking also the observed \oishort\ emission.
The \lws\ subtends a solid angle $\Omega_{\rm LWS}=9.0\times 10^{-8}$\,sr. Disregarding for
a moment the peak emission (see below), we find for unit beam
filling the line intensities $I_{157}= 2 \times 10^{-5}$\,\ecssr\ and
$I_{63}= 1 \times 10^{-5}$\,\ecssr, respectively. Hence, the line ratio \oishort/\ctwo\
is about 0.5 (Fig.\,\ref{oicii_ratio}). These data are consistent with \pdr-emission,
where an interstellar radiation field about 10 times as intense as that of the solar
neighbourhood, i.e. $G_0 \sim 10$, is impinging on the outer layers of a cloud with densities
in the range \nh\,=\,$(0.1 - 1)\times 10^5$\,\cmthree\ (cf. Figs.\,4 and 5 of Liseau et al. 1999).
This estimate of the strength of the UV field is in reasonable agreement with the \fir-background
measured by \iras\ and \iso\ (Larsson et al. 2000), which would imply $G_0 = 5-25$.
In the advocated \pdr\ model, the \oilong\ line is fainter by two orders of magnitude than the
\oishort\ line. The observed line ratio, \oishort/\oilong\,$>$\,12, is clearly consistent
with this prediction. Finally, no detectable emission from higher ionisation stages would be
expected. The \pdr\ model would offer therefore a satisfactory explanation for the observed
fine structure line distribution over the map.

If this \pdr\ emission is treated as a large scale background
and subtracted from the maps, the resulting line ratio toward the peak (SMM\,1) increases dramatically,
viz. to \oishort/\ctwo\,$=15 \pm 5$. Such large ratios are generally not predicted by \pdr\ models
but are a common feature of J-shocks (Hollenbach \& McKee 1989). The residual \oishort\
flux corresponds to an observed intensity $I_{63}= 1 \times 10^{-4}$\,\ecssr, more than two
orders of magnitude below that of J-shock models (see Sect.\,4.1.3). Interpreted as a beam filling
effect, this would imply the size of the shocked regions to be about 4\asec\ to 5\asec.

\subsubsection{The \oishort\ emission toward HH\,460}

Toward the interstellar shock, HH\,460, the \oishort\ flux is not conspicuously larger
than that of \ctwo, as one might naively expect for shock excitation, and we cannot
exclude the possibility that the spatial coincidence with the \oishort\ emission spot
is merely accidental. However, pursuing the shock idea we find that,
for the previously inferred cloud densities, \gapprox\,\powten{5}\,\cmthree,
the \oishort\ intensity is roughly constant with the shock speed
(a few times \powten{-2}\,\ecssr; Hollenbach \& McKee 1989).
These J-shock models do also predict that the accompanying \oilong\ emission
would not be detectable in our observations and that any \ctwo\ contribution would be
totally insignificant.

The observed line intensity is $1\times 10^{-5}$\,\ecssr\ which,
if due to shock excitation, would indicate that the source fills merely a tiny
fraction of the \lws-beam (beam dilution of $2.5\times 10^{-4}$). A size of about 1\asec\
for the \oishort\ emitting regions of the HH object would thus be implied,
which is comparable to the dimension of the dominating, point-like, optical knot HH\,460\,A.
From the observed line flux, a current mass loss rate from the HH-exciting source
of \mdot$_{\rm loss} = 3\times 10^{-7}$\,\msunyr\ would be indicated (Hollenbach 1985,
Liseau et al. 1997), which is at the 3\% level of the mass accretion rate in
the \scc\ (Sect.\,4.4).

Based on the $L_{63} - L_{\rm bol}$ calibration by Liseau et al. (1997),
one would predict that the luminosity of the central source is slightly less than 0.5\,\lsun.
No detailed information about the exciting source of HH\,460 is available, though.
Based entirely on morphological arguments, Ziener \& Eisl\"offel (1999) associate HH\,460 with
SMM\,1, and Davis et al. (1999) either with SMM\,1 or with SMM\,9/S\,68N. The inferred luminosities
of these objects, 71\,\lsun\ and 16\,\lsun, respectively (Larsson et al. 2000), are however
much larger than that inferred for the putative source driving HH\,460. Evidently,
the present status regarding the identification of the driving source of HH\,460 is inconclusive.
Proper motion and radial velocity data would be helpful in this context.

\subsection{\molh, CO, \water\ and OH toward SMM\,1}

We can directly dismiss the \pdr\ of Sect.\,4.1.2 as responsible for the
molecular line emission observed with the \lws, since gas densities and
kinetic temperatures are far too low for any significant excitation of these transitions.
Shock excitation would be an obvious option.
In the following, we will examine the line spectra of \molh, CO, \water\ and OH.

\subsubsection{Rotation diagram: \molh\ and CO}

The analytical technique known as `rotation diagram' analysis
is relatively simple and easy to apply to wavelength integrated molecular rotational
line data. The advantages and the shortcomings of this analysis tool have been
thoroughly discussed by Goldsmith \& Langer (1999).

\begin{figure}
  \resizebox{\hsize}{!}{
  \rotatebox{00}{\includegraphics{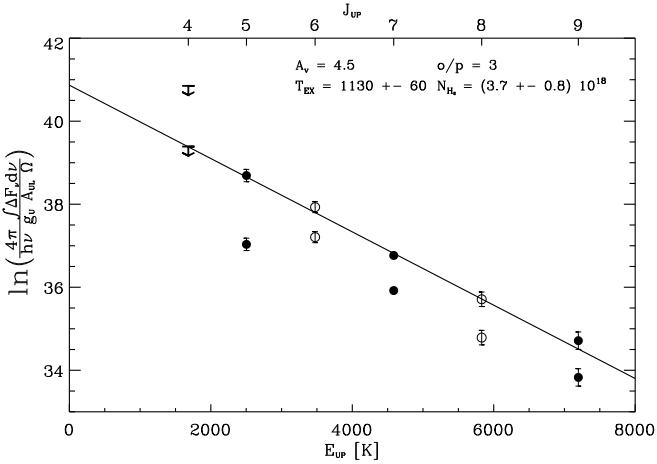}}
                        }
  \caption{Rotation diagram for the ortho-\molh\ (filled symbols) and para-\molh\ (open symbols and
  upper limit) lines observed toward the flow from SMM\,1. Linear regression fits to extinction
  corrected data is shown by the full drawn line. The physical parameters with their formal errors
  are given in the figure (see also the text).}
  \label{h2_rot}
\end{figure}

\begin{figure}
  \resizebox{\hsize}{!}{
  \rotatebox{00}{\includegraphics{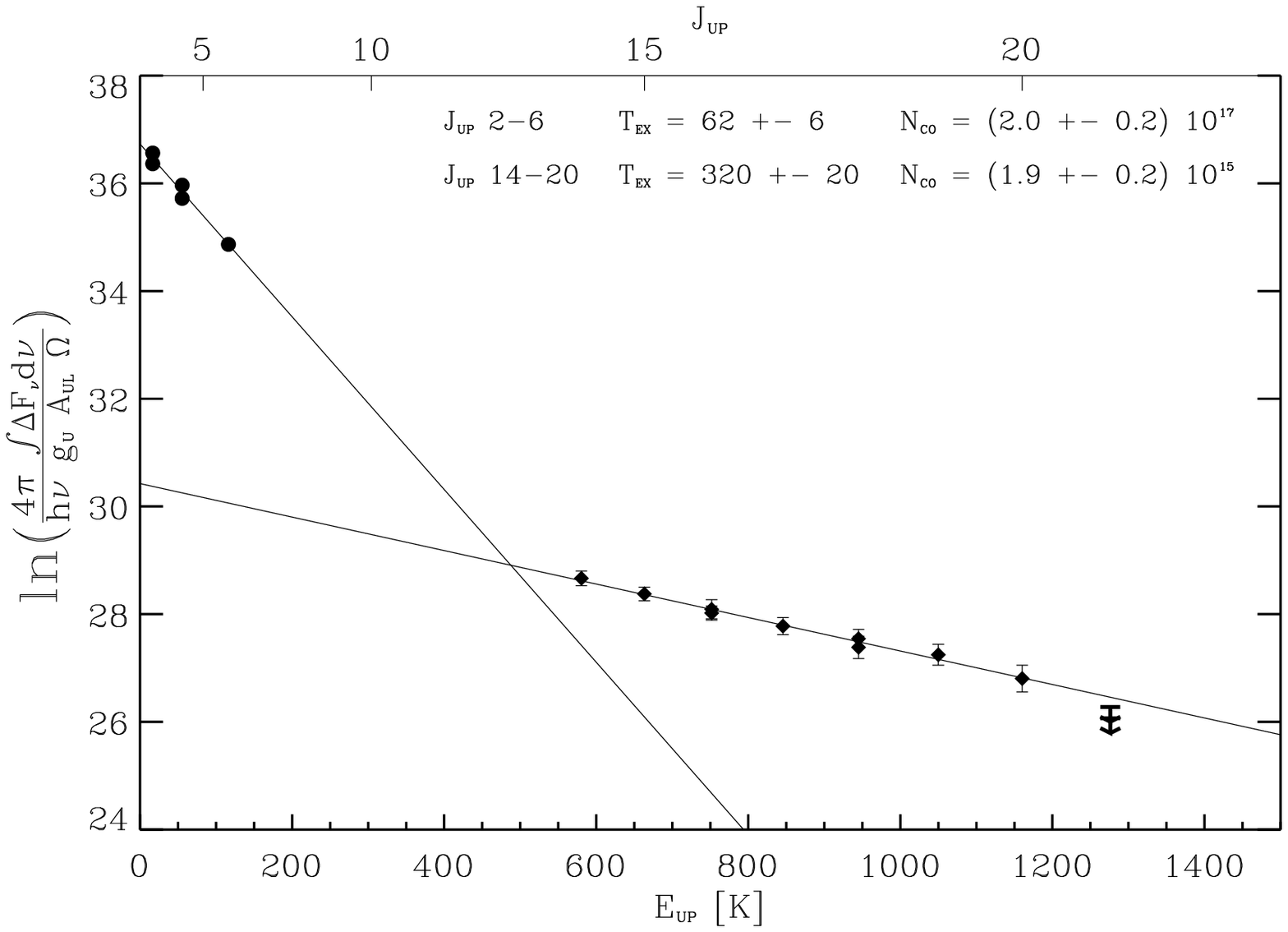}}
                        }
  \caption{Rotation diagram for the CO lines observed with the \lws\ (diamonds, this work) and for
  ground based data (filled circles), taken from Davis et al. (1999), Hogerheijde et al. (1999)
  and White et al. (1995). More than one \lws-value for the same upper energy $E_{\rm u}/k$ refer
  to measurements with different detectors. The data have been (ad hoc) fit by two linear segments,
  the solutions of which are given in the figure (see also the text).}
  \label{co_rot}
\end{figure}

Assuming the lines to be optically thin and to be formed in Local Thermodynamic Equilibrium
(\lte), one can derive the equation of a straight line for the molecular column density as a function
of the upper level energy in temperature units. The slope of this line is the reciprocal
excitation temperature of the levels (which  in \lte\ is the same for all levels
and equals the kinetic gas temperature), viz.
\begin{equation}
\ln{ \left ( \frac { 4\,\pi }{ h\,\nu_0\,g_{\rm u}\,A_{\rm ul} } \,
             \frac{ \int_{\rm line}{F_{\lambda}\,{\rm d}\,\lambda} }{ \Omega_{\rm beam} }   \right )}
=
\ln{ \left ( \frac {N_{\rm mol}}{Q(T_{\rm ex})} \right )}
-
\frac{ E_{\rm u} }{ k\,T_{\rm ex} }
\end{equation}
where the symbols have their usual meaning. The left hand side of Eq.\,(1)
entails the column density of the molecules in the upper levels.
For \molh, the upper level energies, $E_{\rm u}$, were obtained from
Abgrall \& Roueff (1989) and the Einstein transition probabilities, $A_{\rm ul}$,
were adopted from Wolniewicz et al. (1998).
For CO, these data were taken from Chandra et al. (1996). The statistical
weights of the upper levels are given by $g_{\rm u} = (2\,I + 1)(2\,J_{\rm u} + 1)$,
where $I$ is the quantum number of the nuclear spin.
Further, for the evaluation of the approximate partition function
\begin{equation}
Q(T_{\rm ex}) \approx \frac { k\,T_{\rm ex}}{ h\,c\,B_0}
\end{equation}
we used the rotational constant, $B_0 = 59.33451\,{\rm cm}^{-1}$, for \molh\ from Bragg et al. (1982).
For CO, $B_0 = 1.9225\,{\rm cm}^{-1}$, was obtained from the data by Lovas et al. (1979).
Graphs of Eq.\,(1) are shown in Figs.\,\ref{h2_rot} and \ref{co_rot}, fitted to the \cam-\cvf\
and \lws\ data, respectively.

To obtain a consistent result, the \molh\ data need to be corrected for the foreground extinction.
Using the data of Ossenkopf \& Henning (1994; model for thin ice coating, $n =10^5$\,\cmthree, $t=10^5$\,yr),
an extinction correction of \av\,=\,4.5\,mag
resulted in a total column density of warm \molh\ gas of $N({\rm H}_2) = (9.5 \pm 2.0) \times 10^{18}$\,\cmtwo.
The rotation temperature is $T_{\rm ex}=(1060 \pm 100)$\,K and an ortho-to-para ratio (nuclear spin state
population) of $o/p = 3$ is implied by these data.

In Fig.\,\ref{co_rot}, ground-based CO data from the literature were added for lower lying transitions.
Evidently, the high-$J$
distribution appears markedly different from that of the low-$J$ lines. If these latter lines
were truly optically thin, they could originate in extended cloud gas, where
$T_{{\rm low-}J} = (62 \pm 6)$\,K, of column density
$N_{{\rm low-}J} = (2.0 \pm 0.2)\times 10^{17}$\,\cmtwo.
Seemingly in contrast, the \lws\ data identify gas at
a characteristic temperature of $T_{{\rm high-}J} = (320 \pm 20)$\,K, with an
\lte-column density of $N_{{\rm high-}J} = (1.9 \pm 0.2)\times 10^{15}$\,\cmtwo.

These results are based on ad hoc assumptions, i.e. that of unit beam filling and of
low optical depth in the lines, potentially underestimating the column densities,
and that the level populations are distributed according to their \lte\ values. \lte\ may be
a reasonably good assumption for the low-$J$ lines. Regarding CO, it is however questionable to what extent
these are optically thin. On the other hand, low opacity may come close to the truth for
the high-$J$ lines, but \lte\ is not at all guaranteed a priori for these transitions.
Obviously, one needs to check how well these assumptions are justified. In the next sections,
this will be addressed by employing first a method based on the Sobolev approximation
and then a full Monte Carlo calculation, including gradients for both density and temperature.
The latter method takes any (previously neglected) beam dilution effects directly into account.

\subsubsection{Large Velocity Gradient models: CO}

In the Large Velocity Gradient model (\lvg) opacity effects in the lines are explicitly taken
into account by introducing the photon escape probability formalism. Crudely speaking,
the critical density of the transition, $n_{\rm crit} = A_{\rm ul}/C_{\rm ul}(T)$, can be lowered
by means of an effective Einstein-probability, $A_{\rm ul}\,\beta_{\rm esc}$, where $\beta_{\rm esc}$
is in the range 0 to 1 for infinite and zero optical depth, respectively.\footnote{The collision rate
coefficients for CO, $C_{\rm ij}$, have been re-evaluated and extended to higher $J$ values in Appendix\,B.}
This can effectively `delay' line saturation. For illustrating purposes, $\beta_{\rm esc} \sim 1/\tau_{\rm line}$
but, in general, $\beta_{\rm esc}$ is geometry dependent (Castor 1970).

\begin{figure}
  \resizebox{\hsize}{!}{
  \rotatebox{00}{\includegraphics{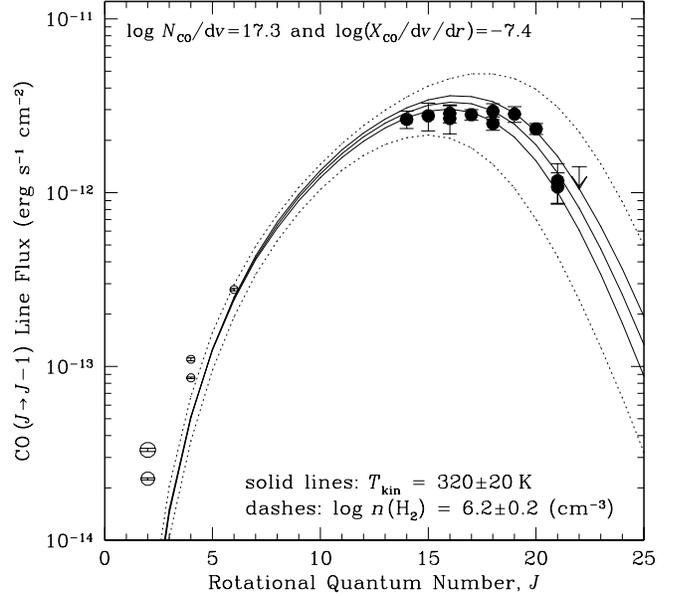}}
                        }
  \caption{The fluxes of rotational lines of CO, observed with the \lws, are compared to \lvg\ model
  computations. Filled circles with error bars refer to \lws\ data, where more than one value for a
  given $J$ correspond to different detectors. The upper limit is $3\sigma$.
  In addition, open symbols represent ground based data,
  where the different sizes refer to different telecope beams (see Fig.\,\ref{co_rot}).
  The physical parameters of the \lvg\ model are indicated in the figure.}
  \label{lvgmod}
\end{figure}

For under-resolved sources, an ambiguity can arise from the fact that hot and tenuous models
may be indistinguishable from cool and dense ones.
However, assuming that the rotation diagram analysis can provide an estimate
of the kinetic gas temperature, \lvg\ can be used to determine the average density of the emitting
region. This is shown in Fig.\,\ref{lvgmod}, for the resulting $\log{n({\rm H}_2)} = 6.2 \pm 0.2$
\cmthree, which is in good agreement with the results by McMullin et al. (2000).

In these models, the presence of a diffuse radiation field is introduced by
the dust temperature $T_{\rm dust} = 40$\,K, the wavelength of unit optical depth
$\lambda_{\tau = 1} =200$\,\um, the frequency dependence of the dust emissivity $\beta = -1$ and
a geometrical covering factor of 0.5 (cf. Larsson et al. 2000). The (clearly detected)
high-$J$ lines are all only mildly sub-thermally excited (justifying a posteriori our initial
assumption), but have substantial opacity, e.g. $\tau_{0(J=14-13)} = 1.7$.
First at $J_{\rm u}=22$ start the lines to become optically thin again ($\tau_{0(J=22-21)} = 0.14$).

The principle parameter of the \lvg\ model is related to the ratio of the column density to the
line width, $N/\Delta v$. For a given density of the collision partners, \nhtwo, this ratio is given by

\begin{equation}
\frac {N_{\rm CO}}{\Delta v} \propto  X_{\rm CO}/\frac{\partial\,v}{\partial\,r}
\end{equation}

where the right hand side is the `\lvg-parameter'. There,
$\partial\,v/\partial\,r$ is the (Doppler) velocity gradient in the gas and $X_{\rm CO}$ is the
molecular abundance relative to \molh. From Eq.\,(3), it is clear that \lvg\ models are, in general,
not unique, since an increase of the column density could have the same effect as a decrease
of the line width.

From the model fit, $N({\rm CO}) = 1.5 \times 10^{18}$\,\cmtwo\ for
the adopted $\Delta v = 7.5$\,\kms\ (\fwhm\ of a Gaussian line shape; see Sect.\,4.3).
A circular source would have a diameter of about 5\asec\ (1500\,AU),
a thickness of about 600\,AU and an \molh\ mass of about 0.01\,\msun\ (for $X_{\rm CO}$\,=\,\powten{-4}).
Finally, the total CO cooling rate amounts to $3.6\times 10^{-1}$\,\lsun.

The hot regions emitting in the \molh\ lines (Sect.\,4.2.1) are not expected to contribute significantly to the CO emission
observed with the \lws. We predict the strongest CO lines from this gas to be the ($J=4-3$)
and the ($J=5-4$) transitions, with ``\lws''-fluxes from a 10\asec\ source of about $1 \times 10^{-14}$\,\ecs.

\begin{figure}
  \resizebox{\hsize}{!}{
  \rotatebox{00}{\includegraphics{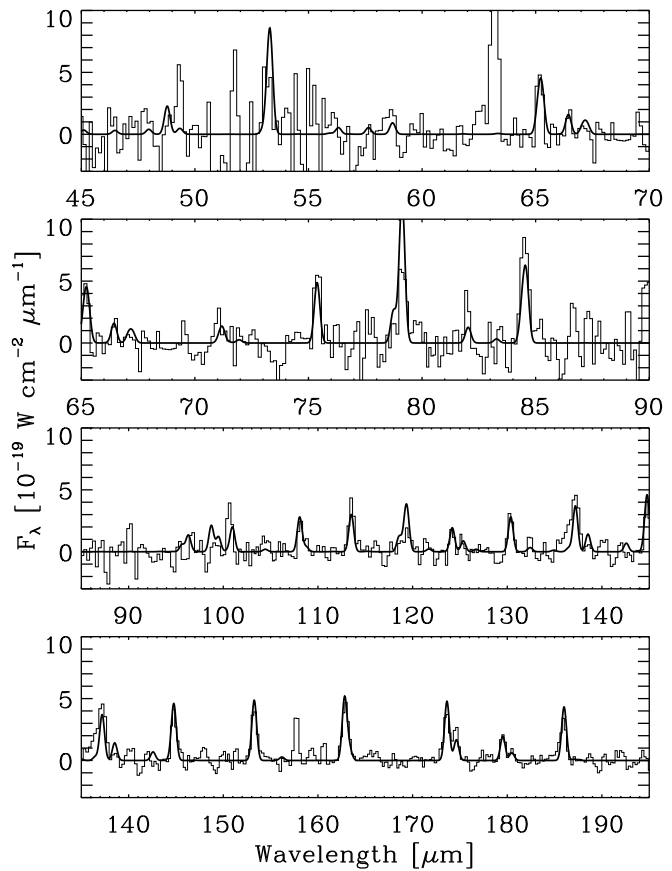}}
                       }
  \caption{The fit (smooth line) of the \lvg\ model to the observed spectrum (histogram). All molecules,
  i.e. CO, $^{13}$CO, ortho-\water, para-\water\ and OH, are assumed to share the same density and temperature, viz.
  $n({\rm H}_2) = 1.6\times 10^6$\,\cmthree, $T_{\rm kin} = 320$\,K; for further model details, see the
  text. Line identifications are as those given in Fig.\,\ref{smm1_spec}.}
  \label{lvg_lines}
\end{figure}

\subsubsection{\lvg\ models: $^{13}$CO, \water\ and OH}

The CO-model of the previous section can be used (by keeping $\partial\,v/\partial\,r$ constant)
to investigate whether it is applicable also to other molecular species. Such a
`one-size-fits-all' model would have the advantage of
permitting the straightforward estimation of the relative abundance of these species
(see Liseau et al. 1996b for an outline of this method).
The reasonably satisfactory result of such computations for $^{13}$CO, ortho-\water, para-\water\
and OH is presented in Fig.\,\ref{lvg_lines}.

The $^{13}$CO spectrum has been computed under the assumption that $^{12}$CO/$^{13}$CO is
as low as 40 (Leung \& Liszt 1976). The data are clearly consistent with this value,
but the S/N is insufficient to conclusively provide a better defined value. Since the $^{13}$CO
lines are all optically thin, the cooling in these lines ($1.3\times 10^{-2}$\,\lsun)
is relatively more efficient than that in CO (by almost a factor of two).

The \water\ model is based on considering 45 levels for both ortho- and para-\water, including
164 transitions each. The radiative rates are from Chandra et al. (1984) and the scaled collision
rates from Green et al. (1993). The model fit of the observed spectrum requires an $o/p = 3$
for \water\ and the derived \water-abundance is $X$(\water)\,=\,$1\times$\powten{-5}. As expected,
the excitation is sub-thermal and the lines are very optically thick (e.g.,
$\tau_{0({\rm o}\,179.5\,\mu {\rm m})} = 433$, $\tau_{0({\rm p}\,101\,\mu {\rm m})} = 218$).
Both the 380\,GHz ortho-transition ($4_{14}-3_{21}$) and the 183\,GHz para-transition
($3_{13}-2_{20}$) are predicted to be masing ($\tau_0 =-1$). The total cooling rate due to water vapour
is $L({\rm H_2O}) = 2.1\times 10^{-1}$\,\lsun, i.e. at the 60\% level compared to the CO cooling rate.

For OH, the Einstein $A$ values were computed from the data provided by D.\,Schwenke
\footnote{Available at http://george.arc.nasa.gov:80/$\sim$dschwenke/}, who also gives the energy levels. The
collision rate coefficients for 50 transitions were obtained from Offer et al. (1994).
Since rates are available for collisions with both ortho-\molh\ and para-\molh, the total rate was based on
\molh-$o/p = 1$. As before, the excitation is sub-thermal
and the lines are optically thick (e.g., $\tau_{0(119\,\mu {\rm m})} \sim 170$ in each line of the doublet).
This refers to the derived, relatively high, value of the OH-abundance of $X({\rm OH}) = 2\times 10^{-6}$
(OH/\water=0.2). The OH lines cool the gas as efficiently as \water, viz. $L({\rm OH}) = 2.0\times 10^{-1}$\,\lsun.
The model is overpredicting
the 119\,\um\ line flux (whereas the 113\,\um\ \water\ line is underpredicted), perhaps indicating
a distribution of temperatures (and densities). However, these
lines fall in one of the least well performing \lws\ detectors (LW\,2) and instrumental effects
cannot be excluded.

Anyway, we have so far considered only models of a homogeneous source at a single kinetic
temperature in a plane-parallel geometry. The relaxation of these, likely unrealistic, assumptions
is the topic of the next sections.

\begin{table}[tbp]
\caption{Main input parameters of the dusty torus model}
\label{modelparams}
\begin{flushleft}
\begin{tabular}{lccl}
\hline\noalign{\smallskip} Parameter&Value\\
\noalign{\smallskip}
\hline\noalign{\smallskip}
\noalign{\smallskip}
Distance                           & 310\,pc\\
Central source luminosity          & 140\,${\rm L_{\odot}}$\\
Stellar effective temperature      & 5000\,K\\
Torus opening angle                & 80{\adeg}\\
Viewing angle                      & 31{\adeg}\\
Torus dust melting radius          & 2\,AU\\
Torus outer boundary               & 1.4\,10$^4$\,AU\\
Torus total mass (gas+dust)        & 33\,${\rm M_{\odot}}$\\
Density at melting radius          & 2.5\,10$^{-13}$\,g\,cm$^{-3}$\\
Density at outer boundary          & 1.4\,10$^{-18}$\,g\,cm$^{-3}$\\
Outflow visual $\tau_{\rm v}$      & 71\\
Midplane $\tau_{\rm v}$            & 2200\\
\hline
\noalign{\smallskip}
\end{tabular}
\end{flushleft}
\label{ModelParams}
\end{table}

\begin{figure}
\begin{center}
  \resizebox{0.75\hsize}{!}{
  \rotatebox{00}{\includegraphics{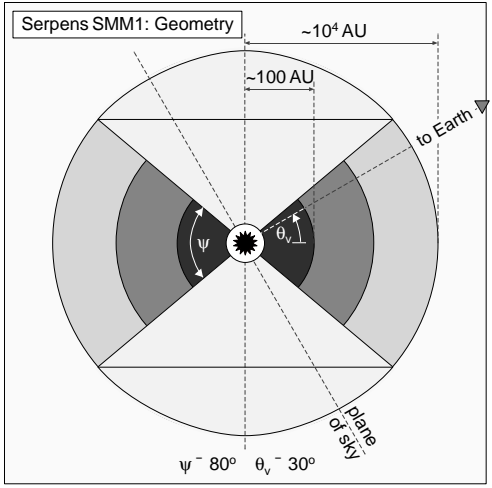}}
                       }
  \caption{Model geometry of the dusty torus of SMM\,1 (see Sects.\,4.2.4
  and 4.2.5). Different shades of gray show schematically the density falling
  off outwards. The radius of the compact dense torus is $\sim$100\,AU, whereas
  the outer radius of the envelope is based on our maps.}
  \label{smm1_geometry}
\end{center}
\end{figure}

\begin{figure}
  \resizebox{\hsize}{!}{
  \rotatebox{00}{\includegraphics{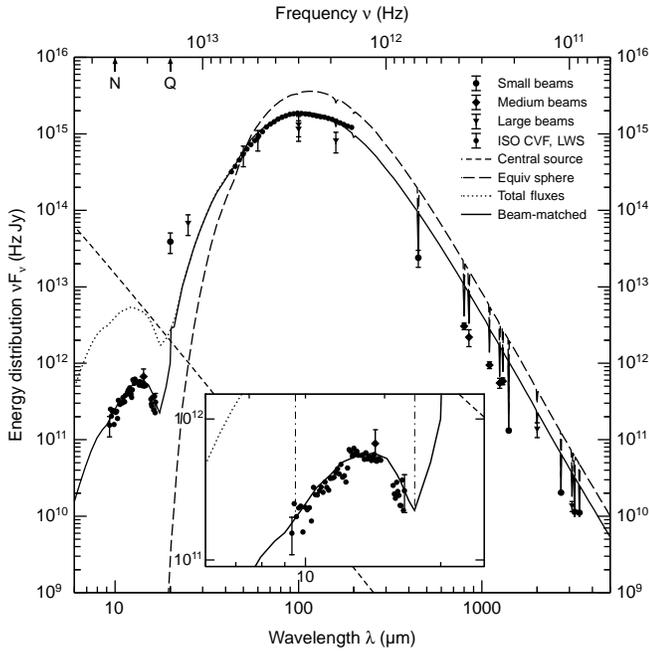}}
                       }
  \caption{Comparison of the observed SED of SMM\,1 and the model of the dusty
  torus. The individual fluxes (see Larsson et al. 2000 for details) are
labelled by different symbols, to distinguish between beams of different sizes.
The model assumes that we observe the torus at an angle of 31{\adeg}, relative
to its midplane. The effect of beam sizes is shown by the vertical lines and
by the difference between the dotted and solid lines in the model SED at mid-IR
wavelengths. Whereas only the lower points of the vertical lines are relevant,
we have connected them to the adjacent continuum by straight lines, to better
visualise the effect. The SED for the equivalent spherical envelope is also
shown, to illustrate the influence of the bipolar outflow cavities.
  }
  \label{smm1_sed}
\end{figure}

\subsubsection{2D radiative transfer model of the dusty torus}

In our previous paper, we presented a self consistent radiative transfer model
for the SED of SMM\,1 (Larsson et al. 2000). For a simplified analysis and for
a direct comparison with previous spherical models of the object, we adopted
spherical geometry of the dusty envelope. The model provided a good fit to the
observations longward of about 60\,\um, but resulted in too low fluxes in the
mid-IR. As already noted in that paper, the spherical symmetry may not be a
very good assumption for SMM\,1, the source driving the bipolar outflow. In
this paper, we performed detailed modelling of the dusty object using our 2D
radiative transfer code (Men'shchikov \& Henning 1997), which enabled us to
quantitatively interpret existing dust continuum observations and to derive
accurate physical parameters of SMM\,1. In the next section, the density and
temperature structure of the model will be used in a Monte-Carlo calculation
of the CO line radiation transfer in the envelope. Our approach and the model
geometry are very similar to those for two other embedded protostars: HL Tau
(Men'shchikov et al. 1999) and L1551 IRS\,5 (White et al. 2000); we refer to
the papers for more details on the general assumptions, computational aspects,
and uncertainties of the modelling.

The model assumes that SMM\,1 consists of an axially-symmetric
(quasi-toroidal), dense inner core surrounded by a similarly-shaped `envelope'
(Fig.\,\ref{smm1_geometry}). A biconical region of much lower density with a
full opening angle of 100\adeg\ is presumed to be excavated in the otherwise
spherical envelope by the outflow from SMM\,1. The structure, for brevity
called 'torus', is viewed at an inclination of 30\adeg\ with respect to the
equatorial plane of the torus. Main input model parameters are summarised in
Table\,\ref{ModelParams}.

\begin{figure*}
\begin{center}
  \resizebox{0.8\hsize}{!}{
  \rotatebox{00}{\includegraphics{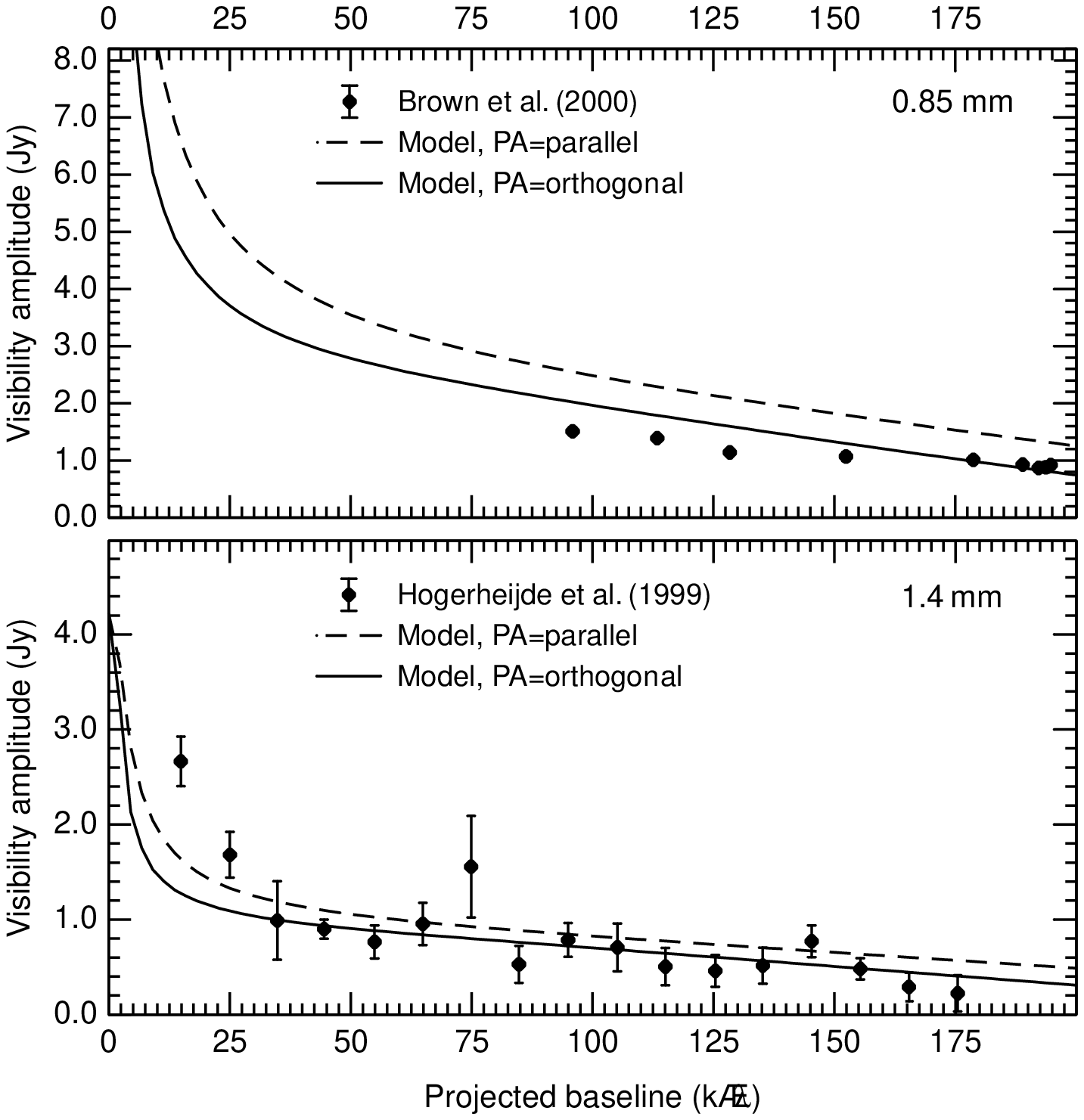}}
  \hspace{-3.5mm}
  \rotatebox{00}{\includegraphics{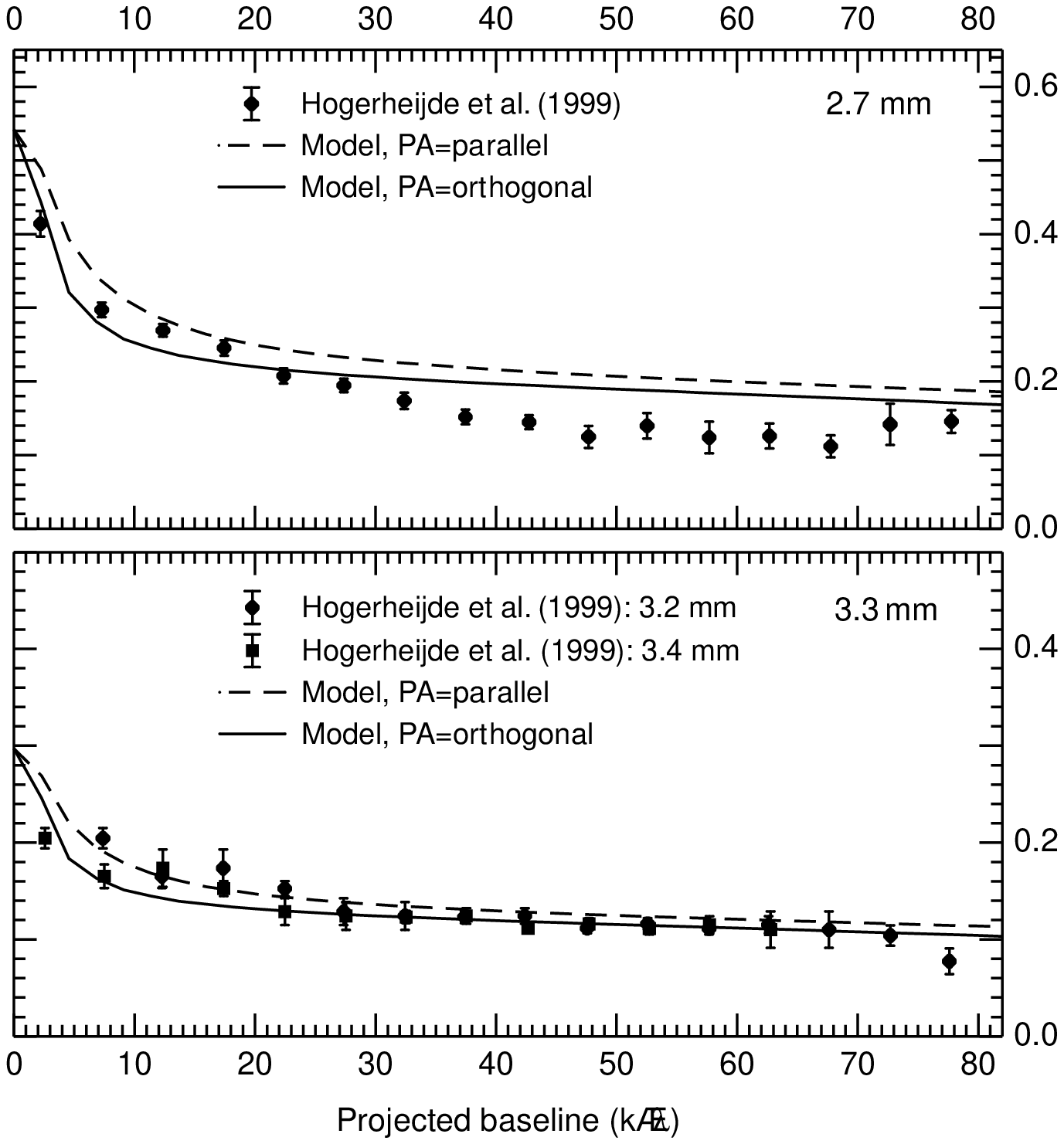}}
                       }
  \caption{Comparison of the model visibilities at 0.8\,mm, 1.4\,mm, 2.7\,mm,
  and 3.3\,mm with available measurements of Brown et al. (2000) and
  Hogerheijde et al. (1999). The upper and lower curves in each panel show
  the visibilities for two directions in the plane of sky, parallel and
  orthogonal to the projected axis of the model.}
  \label{smm1_visibility}
\end{center}
\end{figure*}

\begin{figure*}
\begin{center}
  \resizebox{0.8\hsize}{!}{
  \rotatebox{00}{\includegraphics{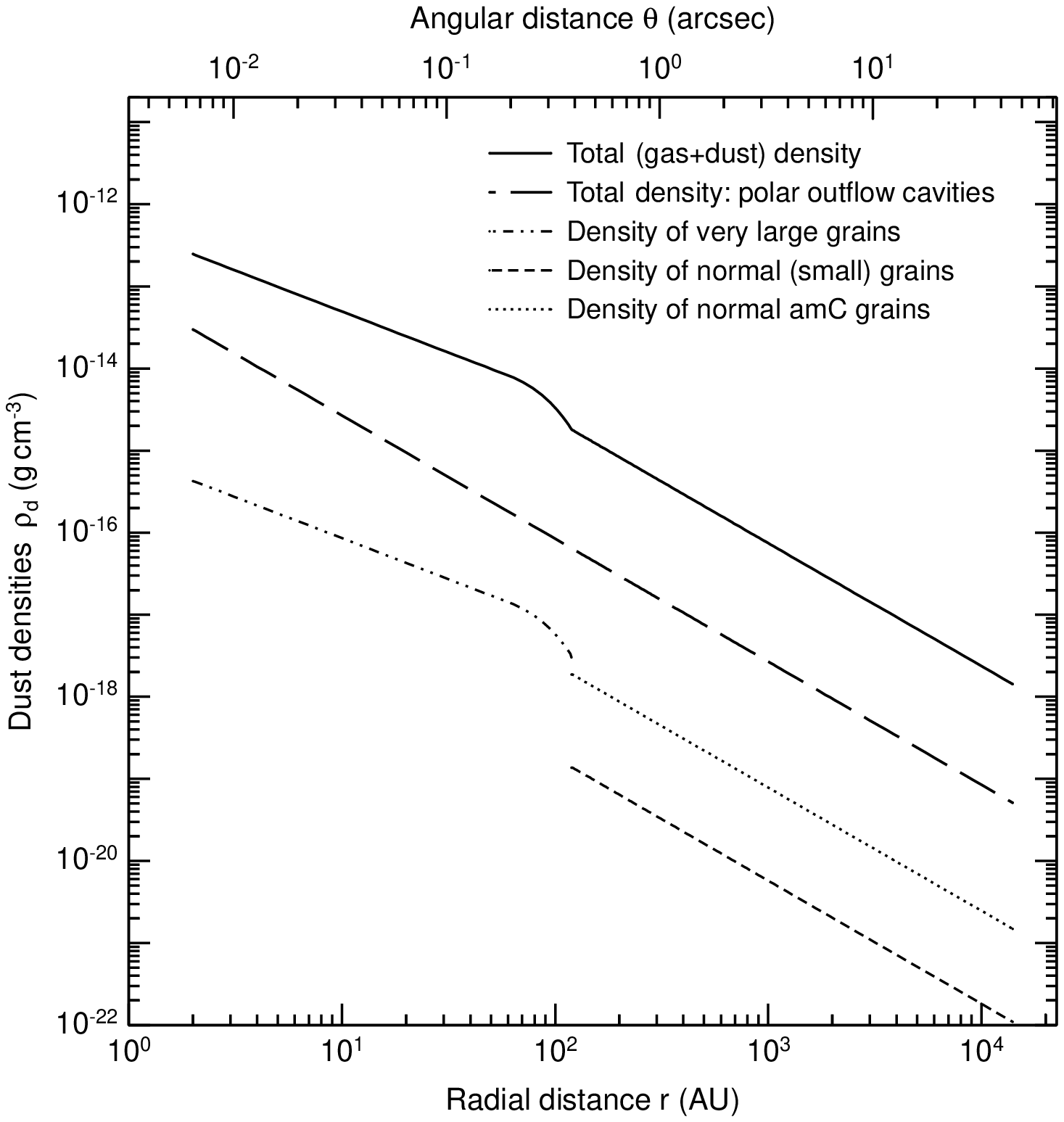}}
  \hspace{-3.5mm}
  \rotatebox{00}{\includegraphics{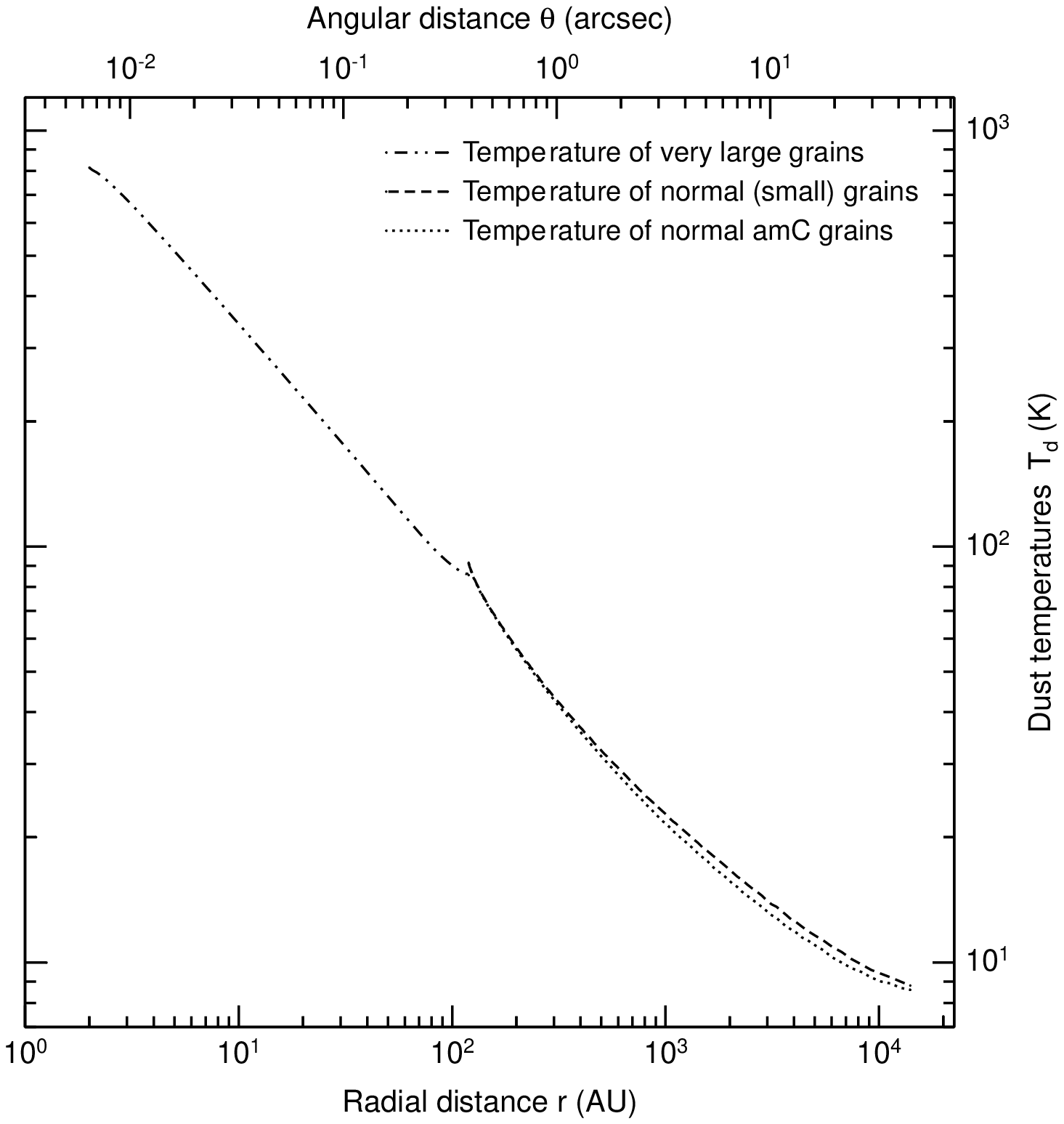}}
                       }
  \caption{Density and temperature structure of the model torus of SMM\,1.
  {\it Left panel:} The total densities in the midplane and in polar outflow regions,
  and dust densities of different dust grain components (for only their smallest sizes).
  {\it Right panel:} The temperature profiles correspond to the midplane of the torus;
  they were obtained self-consistently from the equation of radiative equilibrium.}
  \label{disk_profiles}
\end{center}
\end{figure*}

As very little is known about dust properties in SMM\,1, we adopted a dust
model very similar to that applied by Men'shchikov \& Henning (1999) for HL Tau
and by White et al. (2000) for L1551 IRS\,5. The dust population consists of 4
components: (1) large dust particles of unspecified composition, with radii
100--6000\,\um, (2) core-mantle grains made of silicate cores, covered by dirty
ice mantles, (3) amorphous carbon grains, and (4) magnesium-iron oxide grains.
The latter 3 components of dust grains have the same radii of 0.08--1\,\um. The
dust-to-gas mass ratios of the components are 0.01, 0.0005, 0.0068, and 0.0005,
respectively. The first component of very large grains is present only in the
compact dense torus ($r\,\le\,$120\,AU), where all smaller grains are assumed
to have grown into the large particles. Note that although unknown properties
of dust generally introduce a major uncertainty in the derived model parameters,
extremely high optical depths in SMM\,1 make the model results not very sensitive
to the specific choice of the grain properties, except for the presence of very
large grains in the dense central core.

In the modelling of the dusty torus, we fitted all available photometry of
SMM\,1, paying special attention to the effect of different beam sizes.
Important constraints for the density structure were provided by the available
submm and mm interferometry of the object. The model SED, compared to the
observations in Figs.\,\ref{smm1_sed}, fits almost every single individual flux
in the entire range from the mid-IR to mm wavelengths. Note that it would be
wrong to fit the observed data with the total model fluxes, because the angular
size of SMM\,1 is generally much larger than the photometric apertures. In
fact, the model demonstrates that the effect of beam sizes on the fluxes may be
as large as an order of magnitude.

Comparison of the model visibilities to the interferometry data shown in
Fig.\,\ref{smm1_visibility} demonstrates that the model is also consistent with
the observed spatial distribution of intensity. The visibilities indicate that
there is a dense core inside of a lower density envelope. The radial density
and temperature profiles of the model, are shown in Fig.\,\ref{disk_profiles}.
The innermost dense core has a $\rho\,\propto\,r^{-1}$ density gradient in the
model, whereas the outer parts of the lower-density envelope have a steeper
density distribution ($\rho\,\propto\,r^{-1.5}$). The temperature distribution
was obtained in iterations as a solution of the energy balance equation.

\subsubsection{2D radiative transfer of the molecular emission}

We have used the density and temperature distributions of this dusty torus
model in combination with a Monte Carlo scheme to compute the radiative
transfer of the CO lines, and its isotopomers, through the source.

\begin{figure}
  \resizebox{\hsize}{!}{
  \rotatebox{90}{\includegraphics{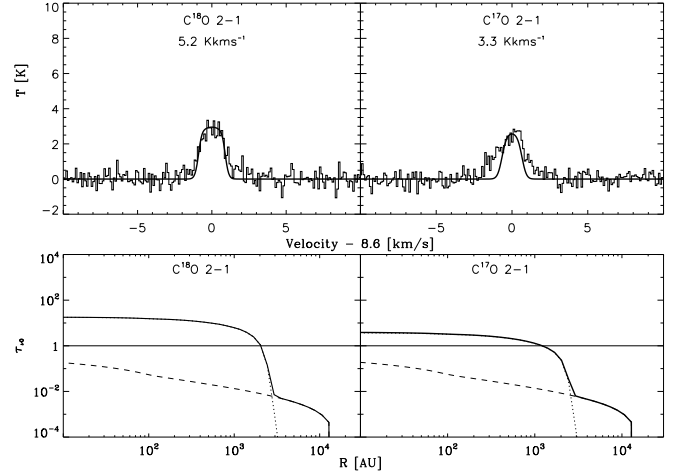}}
                       }
  \caption{{\it Upper panel}: For background emission corrected line profiles of low-$J$ CO isotopomers,
  viz. C$^{18}$O\,($2-1$) and C$^{17}$O\,($2-1$), toward SMM\,1 are shown as histograms. The observations
  were retrieved from the JCMT archive. The results from 2D-Monte Carlo radiative transfer calculations
  for the disk/torus model are shown by smooth lines. The shown integrated line intensity refers
  to the model, for which adopted abundances are
  $^{12}$CO:C$^{18}$O:C$^{17}{\rm O} =  1:550:5$ and freeze-out of the molecules is treated
  in accordance with the work by Sandford \& Allamandola (1993).
  {\it Lower panel}: The total line centre optical depth in C$^{18}$O\,($2-1$) and C$^{17}$O\,($2-1$),
  along the line-of-sight toward the central region of the source, is shown by the solid line. Similarly,
  the dotted lines display the line opacities and the broken lines the dust opacities.}
  \label{mcm_low_j}
\end{figure}

Observations of the \scc\ in the $J=2-1$ transitions of the CO-isotopomers C$^{18}$O and C$^{17}$O
are present in the archive of the James Clerk Maxwell Telescope (JCMT). These potentially
optically thin lines could trace the embedded core SMM\,1. Our disk model reproduces the observed
line intensities of these low-$J$ isotopomers fairly well (Fig.\,\ref{mcm_low_j}). There,
the averaged background emission of the surrounding gas has been subtracted, in order to reveal the
line profiles of SMM\,1 itself. From the figure it is evident that the C$^{18}$O and C$^{17}$O lines
are optically thick out to a point, where the temperature falls below 15\,K and where substantial
condensation of the CO gas onto dust grains occurs. This CO freeze-out was treated following
Sandford \& Allamandola (1993 and references therein), where the ice-to-gas ratio $\eta$ is proportional to
the dust density $n$, and to functions of the gas and dust temperatures, $T_{\rm g}$ and $T_{\rm d}$
respectively, viz.

\begin{equation}
      \eta \propto n\,T_{\rm g}^{1/2}\, e^{1/T_{\rm d}}
\end{equation}

\begin{figure}
  \resizebox{\hsize}{!}{
  \rotatebox{90}{\includegraphics{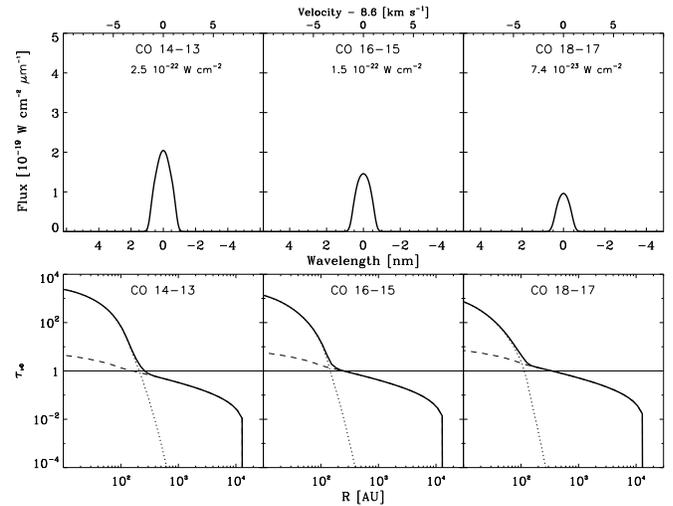}}
                       }
  \caption{Same as in Fig.\,\ref{mcm_low_j}, but for three high-$J$ CO transitions,
  which fall in the ISO-LWS spectral band.}
  \label{mcm_opacity}
\end{figure}

In the inner (`core') regions, line opacities are very high
so that the photons are essentially trapped. Therefore, the excitation temperature of the molecules follows the
kinetic temperature of the \molh\ gas. Beyond the core, i.e. in the `torus', $\tau$ drops quickly below unity
for the high-$J$ lines, but there, the dust starts to contribute significantly to the (small) opacity
(Fig.\,\ref{mcm_opacity}).

The CO lines falling into the \lws\ range are all formed in the inner, hotter parts
of the source ($r \ll 10^3$\,AU, $T \gg 10^2$\,K). This small line forming region is insufficient
to produce the observed flux levels, i.e. the model underpredicts observed high-$J$ line
fluxes by more than two orders of magnitude. Irrespective of the geometry, we can conclude quite generally
that the CO lines observed with the \lws\ do not originate from the central regions of SMM\,1, be it an
accretion disk, be it infalling gas (we have also computed `inside-out' collapse models).

For the excitation of this gas we need to consider alternative mechanisms and, since
outflows are known to exist in this region, shock heating of the gas offers a natural option.
Our temperature determinations for the molecular gas (Sect.\,4.2.1) are also consistent with this idea.

\subsection{Shock heating}

From the discussion of the preceding sections we can conclude that the heating of the gas is
most likely achieved through shocks. These shocks are generated by flows within the \lws\ beam.
Comparing the observed and predicted molecular line emission with the J-shock models
by Hollenbach et al. (1989) and Neufeld \& Hollenbach (1994), we find that these models
are in conflict with our observations.

\begin{figure}[h]
  \resizebox{\hsize}{!}{
  \rotatebox{00}{\includegraphics{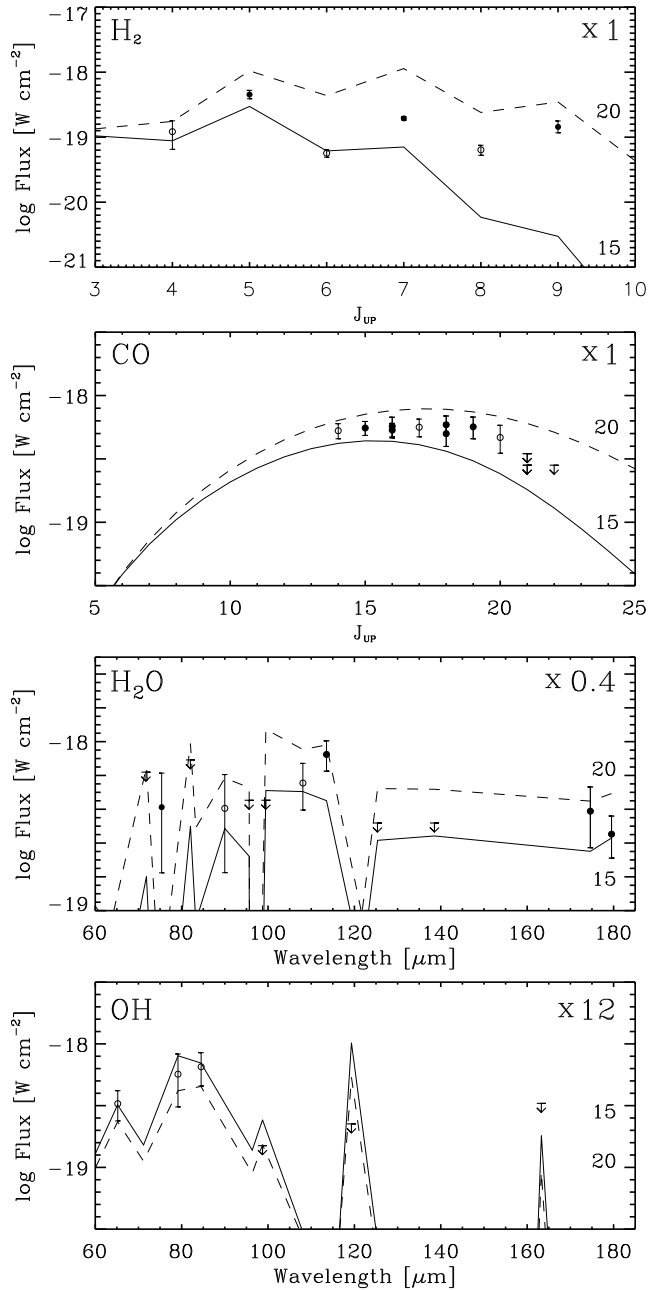}}
                        }
  \caption{Comparison of our molecular line observations with the predictions of theoretical models
  of C-shocks (Kaufman \& Neufeld 1996). The pre-shock density is always $\log n_0 = 5.5$ (\cmthree).
  Solid and dashed lines are for models with \vs\,=\,15\,\kms\ and 20\,\kms, respectively. The \molh\
  models refer to the \cvf\ pixel size, whereas the other panels are for a circular source of
  diameter 11\asec. The scaling factors, necessary to bring the observed and model data into agreement,
  are indicated in each frame by `$\times\,{\rm factor}$'.}
  \label{kn_models}
\end{figure}

In Fig.\,\ref{kn_models}, we compare our observations of rotational lines of \molh, CO, \water\ and OH with
predictions of the C-shock models by Kaufman \& Neufeld (1996). The models for $\log n_0 = 5.5$ (\cmthree)
and \vs\,\about\,15 -- 20\,\kms\ are in reasonable agreement with the for extinction corrected
(\av\,=12\,mag) observed values for \molh\ (Kaufman \& Neufeld used $o/p=3$) and for a flux from
$6^{\prime \prime} \times 6^{\prime \prime}$ (1 \cvf-pixel). For CO, the model fits the observations
for an adopted circular source of diameter 11\asec. To achieve agreement for \water, the model
fluxes would need to be adjusted downwards by a factor of 2.5, whereas an increase by more than one
order of magnitude (a factor of 12) would be required for OH. Evidently, OH is largely underproduced
by these models, a fact also pointed out by Wardle (1999). If on the other hand the Wardle model
is essentially correct, this would suggest that the ionisation rate in the \scc\ is significantly higher
(up to $10^{-15}\,{\rm s}^{-1}$) than on the average in dark clouds,
$\zeta =(10^{-18}-10^{-17})\,{\rm s}^{-1}$.
High X-ray activity is known to be present within the \scc\ (Smith et al. 1999 and references therein).
It is conceivable that such a high ionisation rate could also have considerable consequences
for the cloud chemistry and its evolution. For instance, a relatively higher H$^+_3$ abundance
could be expected, the effects of which (in addition to the enhanced abundance of OH) may in fact
have already been observed (e.g., HCN/HNC\,\about\,1; McMullin et al. 2000 and references therein).

\subsection{Summarising discussion}

Based on their 0.8\,mm JCMT-CSO interferometry, Brown et al. (2000) obtained estimates of the size, mass
and average (dust) temperature of the disk of SMM\,1. The estimated mass is larger and
the size of the disk is smaller by one order of magnitude than what is required
to account for the observed level of line emission (Sect.\,4.2.2).
Unless the disk (extended atmosphere?) is heated to very much higher temperatures
(by an as yet to be identified mechanism) than the 60\,K determined by Brown et al.,
we find it unlikely that the molecular line spectrum of SMM\,1 is of circumstellar
disk origin. Our own calculations (Sect.\,4.2.4) confirm this conclusion.

It is intriguing that the luminosity of the spherical model of SMM\,1 (71\,\lsun, Larsson et al. 2000)
is close to the `magic number' of the classical main accretion phase of solar mass stars (Shu et al. 1987).
At the elevated cloud temperature of the \scc\ (\about\,40\,K, White et al. 1995), the
isothermal sound speed is 0.4\,\kms\ and, hence, the (time averaged, cf. Winkler \& Newman 1980)
mass accretion rate corresponds roughly to \mdot$_{\rm acc} = 10^{-5}$\,\msunyr,
yielding $\,L_{\rm acc} = 70$\,\lsun, where we have used the mass-radius relationship of
Palla \& Stahler (1990). In this scenario, the age of SMM\,1 would be about $10^5$\,yr or less,
depending on the details of the acquired mass of the (presumably deuterium burning) central core.
Regarding the data presented in this paper, we find it however difficult to reconcile this
accretion shock model with our observations. As concluded in Sect.\,4.2.4, the excitation of
the observed lines requires significantly larger volumes at elevated densities and temperatures.

The \molh\ observations are partially resolved and there exists no ambiguity as to where, with respect
to SMM\,1, the emission arises (cf. Fig.\,\ref{h2_map}). These lines trace a collimated outflow
toward the northwest of SMM\,1, which is also
seen in ro-vibrationally excited \molh\ line emission (Eiroa \& Casali 1989; Hodapp 1999). In
the graphs of Fig.\,\ref{kn_models}, we have assumed that also the \lws\ lines originate essentially at the
location of the \molh\ spots (i.e. we have artificially introduced another factor of two for the fluxes).
However, the dereddened data of Eiroa \& Casali (with the \av-value determined in Sect.\,4.2.1)
could potentially present an additional difficulty for the C-shock model (Kaufman \& Neufeld 1996).
The estimated 1-0S(1) line intensity would in this case
be larger by more than two orders of magnitude than that predicted by the model. We cannot exclude
at present, however, the possibility that the 1-0S(1) emission observed by Eiroa \& Casali (1989)
is essentially unextinguished. Photometrically calibrated data at higher spatial resolution would be
required to settle this issue.

The mechanical energy input by the flow is
$L_{\rm mech} = 0.5\,\mu_{\rm gas}\,m_{\rm H}\,n_0\,v_{\rm s}^3 \times {\rm area}$, which for
a pre-shock density of $\log{n_0} = 5.5$, a shock velocity \vs\,=\,($15-20$)\,\kms, and a
5\asec\ source size yields $L_{\rm mech} = (6.4-15.1) \times 10^{-2}$\,\lsun.
From the Kaufman \& Neufeld (1996) C-shock model, this gas is cooled by \molh\ at a rate of
$(1.0-4.7) \times 10^{-2}$\,\lsun. From the \lws\ data, we
inferred the total cooling rate through the lines of CO, $^{13}$CO, \water\, and OH of
$78\times 10^{-2}$\,\lsun\ (Sects.\,4.2.2 and 4.2.3), corresponding to 0.5\% to 1\% of the total
dust luminosity. This is larger by factors of 5 to 12 and it is thus not excluded
that the shocked regions observed in the \molh\ lines and those giving rise to the
\fir\ lines are not the same. We reached the same conclusion on the basis of our
excitation and radiative transfer calculations.

The observed and background-corrected \oishort\ emission toward SMM\,1 suggests a
contribution also by J-shocks within the \lws\ beam (Sect.\,4.1.2). Intriguingly, the
derived dimensions are practically identical to those determined for the \lws-molecular
emission, albeit existing J-shock models do not predict the relative intensities correctly.
At present, we can merely conclude that shocks, in general, provide a plausible energy input mechanism,
although the details of the shock type(s) are less clear. We propose that predominantly slow shock
waves in the dense medium surrounding SMM\,1 provide the heating of the molecules we have
observed with \iso, whereas dynamical collapse is not directly revealed by our data.

\section{Conclusions}

Based on spectrophotometric \iso\ imaging with the \lws\ and the \cam-\cvf\ of significant
parts of the active star forming \scc\ our main conclusions can be summarised as follows:

\begin{itemize}
\item[$\bullet$] We find the emission in the \oishort\ and \ctwo\ fine structure lines to be
extended in our $8^{\prime} \times 8^{\prime}$ map. The absolute intensities and their ratios
can be explained in terms of \pdr\ models, where a UV field of $G_0 = 15 \pm 10$ is falling
onto the outer layers of the dark cloud, where densities are of the order of $(10^4 - 10^5)$\,\cmthree.

\item[$\bullet$] Also the emission in rotational lines of \water\ and high-$J$ CO appears (slightly)
extended, but we cannot exclude the possibility that it arises from point sources in the field,
viz. the Class\,0 objects SMM\,9/S\,68 and SMM\,4. The maximum of the emission is observed toward
SMM\,1, the dominating far infrared and submm source in the \scc.

\item[$\bullet$] The spectrum of SMM\,1 contains numerous lines of CO, \water\ and OH. These lines
are generally subthermally excited and optically thick and trace regions of dimensions \about\,\powten{3}\,AU
($\sqrt{1500 \times 600}$\,AU),
where temperatures are above 300\,K and densities above \powten{6}\,\cmthree.
The derived abundances, relative to \molh, are for CO, \water, OH and $^{13}$CO, respectively,
$X_{\rm mol}=(1,\,0.1,\,0.02,\,\ge 0.025) \times 10^{-4}$. The ortho-to-para ratio for \water\
is consistent with the high temperature equilibrium value (ratio of the statistical weights of the nuclear spins),
i.e. \water-$o/p =3$.

\item[$\bullet$] The relatively high OH abundance is indicative of an elevated level of
ionising flux in the \scc, causing the ionisation rate to be $\zeta \gg 10^{-18}\,{\rm s}^{-1}$,
i.e. significantly higher than the average rate prevailing in dark clouds. Strong and active
X-ray sources, known to exist in the cloud, could be responsible for this radiation.

\item[$\bullet$] The observed SED of SMM\,1 is consistent with a model of a
dusty torus with an outer radius of 14\,000\,AU (45\asec). The torus is heated
by a central stellar source, with a (somewhat arbitrarily) adopted effective
temperature of 5000\,K. A luminosity of 140\,\lsun\ is required to explain the
observations. The total mass of the toroidal core of SMM\,1 is 33\,\msun. The
derived visual extinction through the torus exceeds 2000\,mag and the torus is
optically thick up to mm wavelengths.

\item[$\bullet$] 2D modelling of the radiative transfer through the
circumstellar torus of SMM\,1 reveals that it is highly unlikely that the
observed molecular emission arises in the torus. The same conclusion can be
drawn for models of dynamical collapse (`inside-out' infall).

\item[$\bullet$] The \molh\ data have been obtained at significantly higher spatial resolution
than that offered by the \lws. These \cam-\cvf\ observations trace the regions of pure rotational
\molh\ line emission. The \molh-maxima are observed to be displaced from SMM\,1 and are situated
toward the northwest, along the jet of outflowing material from this Class\,0 source.The temperature
of \about\,\powten{3}\,K of this \molh\ gas is indicative of the heating by relatively slow shock waves. 

\item[$\bullet$] The comparison of our molecular line data with shock models suggests that
shock heating with \vs\,\about\,(15--20)\,\kms\ along the outflow of SMM\,1 is the most likely
mechanism of molecular excitation, although the details of these shocks are less clear.
\end{itemize}

\acknowledgements{We are grateful for the help with the data reductions of the \cvf\ observations
by Stephan Ott. We also thank Ewine van Dishoeck for making avalailable to us the collision
rate coefficients for OH in electronic form. The support of this work by {\it Rymdstyrelsen}
(Swedish National Space Board) is acknowledged.}

\appendix

\section{Reduction of the \lws\ data}

The photometric calibration of the \lws\ is primarily based on observations
and models of the planet Uranus. The S/N in these observations is rather modest,
which will affect the relative spectral response function (RSRF) derived from these data.
When calibrating the `science data', the registered photocurrent is
divided by the RSRF, propagating any uncertainty in the
RSRF, which will ultimately lead to errors in the derived flux density, $F_{\lambda}$.

\begin{figure}[t]
  \resizebox{\hsize}{!}{\rotatebox{00}{\includegraphics{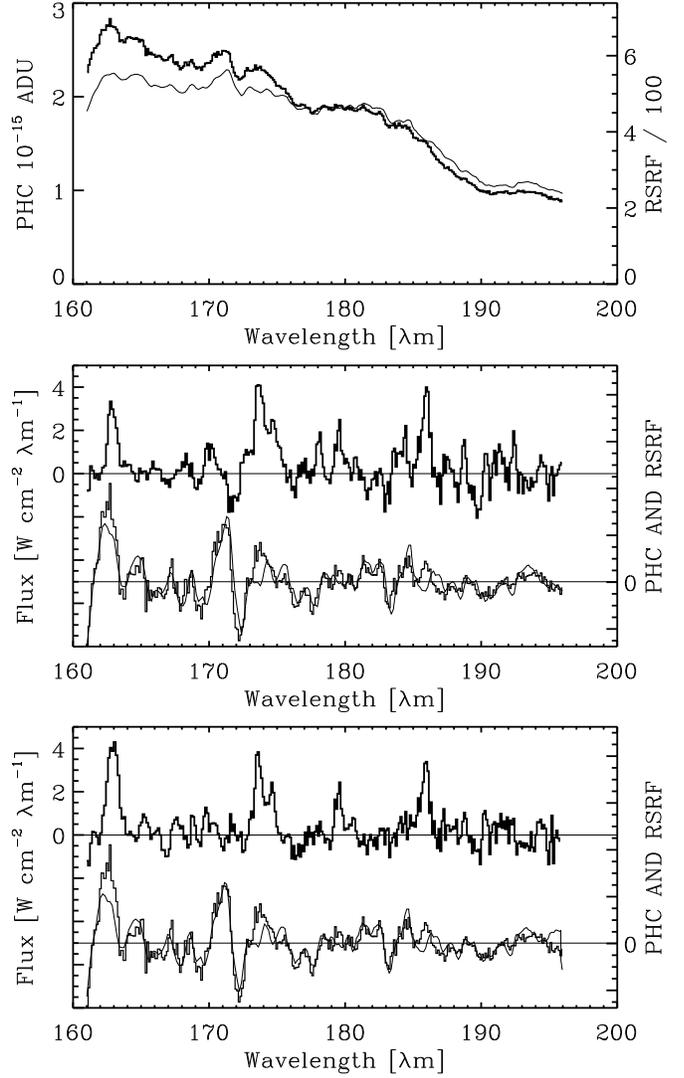}}}
  \caption{{\bf Upper panel:} The observed wavelength dependence of the photocurrent (thick line) 
  for the detector LW\,5 of the \lws. The thin line refers to the grating response
  function (RSRF) of the detector, scaled to the same mean value as the photocurrent.
  {\bf Middle panel:} The lower curves are the same as in the upper panel, after subtracting
  boxcar fits. The division of the photocurrent by the RSRF results in the upper curve, which is the
  observed flux density (for clarity a median fit to the continuum has been subtracted).
  {\bf Lower panel:} The same as in the middle panel, but with a small
  wavelength shift ($< 0.15 \mu$m) of the observed photocurrent, before
  division. Evidently, the `noise' is much reduced: compare, e.g., the spurious
  `absorption' feature near 172\,\um\ in the middle and lower panels.}
  \label{phcresp}
\end{figure}

The derived photocurrent after standard processing, resulting in an SPD file and
where identified instrumental peculiarities have been removed, is shown for the
LW\,5 \lws\ detector in the upper panel of Fig.\,\ref{phcresp}, together with the
relative spectral response function scaled to the same mean value.
From the figure, it is evident that many features seen in the
photocurrent are due to the detector and grating response.

The middle panel of Fig.\,\ref{phcresp} shows again the
photocurrent and the RSRF, but now after the subtraction of a
wide boxcar smoothing function (continuum subtraction).
The upper curve shows  the result of the division
of the photocurrent by the response function. This allows us to gauge
the effects on the spectra from the narrow features of the RSRF.
In the long wavelength regime of the \lws, and in LW\,5 in particular,
there are spurious absorption features at positions, where the
RSRF is steep. In regions, where the RSRF has steep gradients, can already
very small wavelength errors create large features that could be mistaken
for spectral lines. A wavelength mismatch between the photocurrent
and the RSRF will also introduce an overall lowered  S/N.

In the lower panel of Fig.\,\ref{phcresp}, a relative shift by less than a
quarter of an resolution element (the original sampling rate was at four times
the spectral resolution) of the photocurrent resulted in the considerable reduction
of an apparent broad `absorption' feature near 172\,\um, an overall better S/N and
therefore a better definition of the spectral lines.

All individual spectra for all ten detectors have been carefully monitored
for obviously anomalous features which (most likely) were introduced by the RSRFs.
This was done for the RSRFs of both OLP\,8 and OLP\,10 and any such spurious features
were, of course, corrected for. This does not guarantee, however, that no such
false spectral features do still exist in our data, as we were very restrictive
in our application of any wavelength shifts.

\section{Collision rates for CO}

We used rate coefficents for collisions of CO with \molh\ which
are based on values found in the literature but which have been extended
to rotational quantum numbers $J_{\rm u} = 40$, although extrapolations to
higher $J$ is not excluded.

\begin{figure}
  \resizebox{\hsize}{!}{\rotatebox{0}{\includegraphics{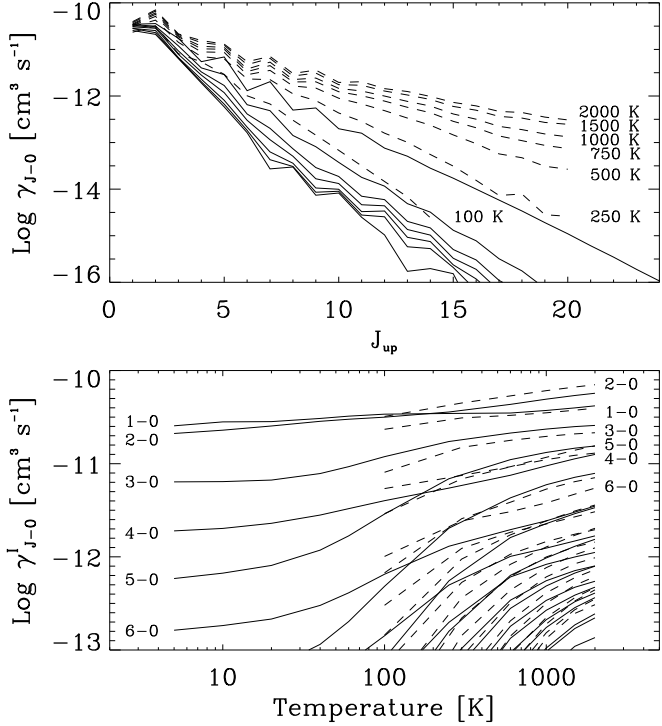}}}
  \caption{Downward rate constants for collisions of CO with para-\molh:
  {\bf Upper panel:} as a function of $J$, where the solid lines are for
  $T = 5,\,10,\,20,\,40,\,60,\,100\,{\rm and}\,250$\,K from Flower (2001),
  and the dashed lines for $T = 100,\,250,\,500,\,750,\,1000,\,1500\,{\rm and}\,2000$\,K
  from Schinke et al. (1985). {\bf Lower panel:} as a function of $T_{\rm kin}$,
  with the data by Schinke et al. adjusted to fit those by Flower at 400\,K,
  expanding the rates for the low temperatures to $T=2000$\,K (see the text).
  These rates are labelled $\gamma^{\rm I}$ in the figure.}
  \label{coll1}
\end{figure}

For $J_{\rm u} \le 29$ and for low temperatures, ($5 - 400$)\,K,
we used the recent rates of Flower (2001; ortho-\molh-CO and para-\molh-CO;
downward rates). For the higher temperature range of ($>400 - 2000$)\,K, the
calculations by Schinke et al. (1985; para-\molh-CO; upward rates) were used.
The matching between these data sets is roughly acceptable, but there
exist disagreements (Fig.\,\ref{coll1}), which reflect the differences in
assumptions and computational methods (see the discussion of resonances by
Flower).

In order to arrive at a consistent set of collision rate constants for
the hole range of temperatures, the Schinke et al. data (correctly
transformed to de-excitation rates; see also: Viscuso \& Chernoff 1988)
were laterally shifted to fit the Flower data at 400\,K. A satisfactory matching
was, however, not really possible for the lowest transitions connecting to
the ground state, see the lower panel of Fig.\,\ref{coll1}.
In the figure, these expanded rates are labelled $\gamma^{\rm I}$.

\begin{figure}
  \resizebox{\hsize}{!}{\rotatebox{0}{\includegraphics{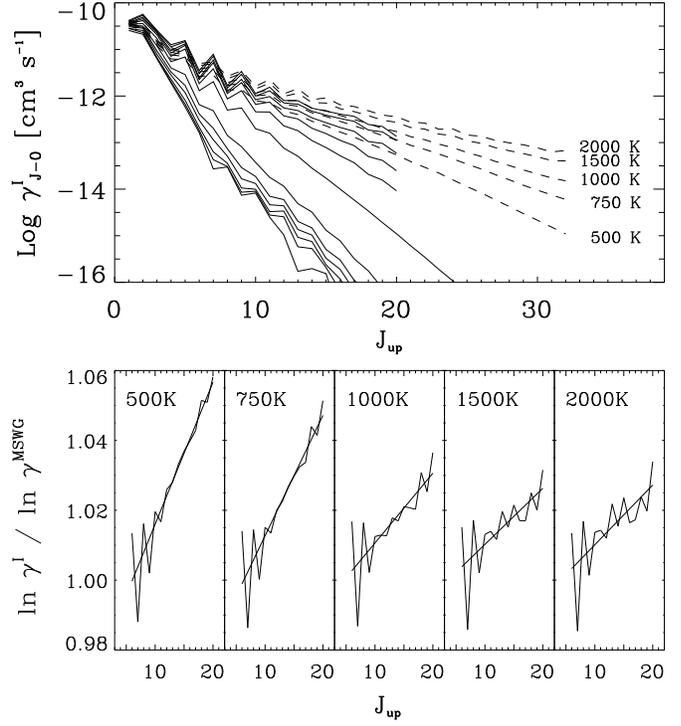}}}
  \caption{Rate constants for CO as a function of $J$:
  {\bf Upper panel:} the solid lines refer to the rescaled data of Fig.\,\ref{coll1},
  i.e. $\gamma^{\rm I}$, whereas the dashes are for
  $T = 500,\,750,\,1000,\,1500\,{\rm and}\,2000$\,K
  from McKee et al. 1982 (collisions with He, scaled to \molh, and up to $J = 32$).
  {\bf Lower panel:} the fitted ratios of these rates ($J_{\rm u} \le 20$) as a
  function of $J_{\rm u}$ and, as an example, for the five selected temperatures.}
  \label{coll2}
\end{figure}

\begin{figure}
  \resizebox{\hsize}{!}{\rotatebox{0}{\includegraphics{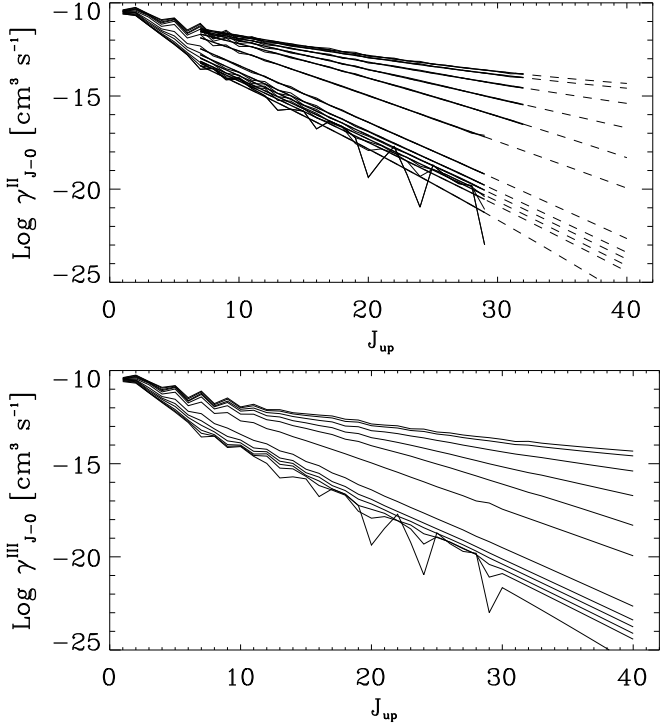}}}
  \caption{{\bf Upper panels:} Applying the curvature corrections to the data by McKee et al.
  results in the rates named $\gamma^{\rm II}$ and shown by the solid lines. The extrapolations
  of $\gamma^{\rm II}$ to high $J$-values are indicated by the dashed lines.
  {\bf Lower panel:} The finally adopted collision rate coefficients for CO,
  $\gamma^{\rm III}$, for the hole range in temperature, $T= (5 - 2000)$\,K, and energy, $J = 0 - 40$.}
  \label{coll3}
\end{figure}

These $\gamma^{\rm I}$ data span $J$-values up to 20 and temperatures between 400 and 2000\,K.
For the same temperature intervall,
McKee et al. (1982, MSWG) have published calculations (He-CO; downward rates)
for $J$-values up to 32  (Fig.\,\ref{coll2} upper panel).
These rates ($\times 1.37$) were divided into the Schinke et al. rates and
fit by a polynomial to correct the shape of the McKee et al. data
(Fig.\,\ref{coll3} upper panel), viz.

\begin{equation}
  \begin{array}{lll}
      \ln{\gamma^{\rm I}}/\ln{\gamma_{_{\rm MSWG}}} = a + b\,J &  &  20 \leq J \leq 32
  \end{array}
\end{equation}
to obtain a new set of rate coefficients

\begin{equation}
      \gamma^{\rm II} = \gamma_{_{\rm MSWG}}^{a + b\,J}
\end{equation}

For the expansion to $J=40$, the $\gamma^{\rm II}$ were fit to a second order polynomial, viz.

\begin{equation}
  \begin{array}{lll}
  \ln{\gamma^{\rm II}}  = c + d\,J + e\,J^2  &  &  7 \leq J \leq ^{29}_{32}
  \end{array}
\end{equation}
the result of which was used for the extrapolations of the para-\molh\ rates up to $J = 40$, viz.

\begin{equation}
  \begin{array}{lll}
    \gamma^{\rm III} = \exp{({c + d\,J + e\,J^2})} & &   J > ^{29}_{32}
  \end{array}
\end{equation}
This provides us finally with the full set of collision rate coefficents for $J_{\rm u} \rightarrow 0$ for
$J_{\rm u}$ up to 40 (Fig.\,\ref{coll3}).

So far, we have considered rates only connecting to the ground state, i.e. $\gamma_{J-0}$. Flower (2001)
did provide rates for collision transitions between all level's, but for higher $J$-values and/or higher
temperatures we don't have that information. If the kinetic energy of the collision partners on the other hand
is large compared to the rotational energy spacing of the CO molecule,
the other rate cofficients can be obtained from (Goldflam et al. 1977; McKee et al. 1982)
\begin{equation}
\begin{array}{lll}
  \gamma_{J-J'} &  =  & \left(2J'+1\right) \times \\
           &     & \times \sum_{L}^{} \left(2L+1\right)
        \left(
                \begin{array}{ccc}
                 J & L & J' \\
                0 & 0 & 0
                \end{array}
                \right)^{2} A_{L,J}^{2}\,\,\gamma_{L-0}
  \end{array}
\end{equation}
where the Wigner 3-$j$ symbol designates the Clebsch-Gordan coefficients, which
were computed exactly (see, e.g., Zare 1986). Further, the factor $A_{L,J}$

\begin{equation}
A_{L,J} = \frac {6 + (a L)^2}{6 + (a J)^2},\,{\rm where}\hspace{0.15cm}
a = 0.13\,B_0\,l_{\rm sca} \left ( \frac {\mu}{T} \right )^{\frac {1}{2}}
\end{equation}
given by McKee et al. (1982), should correct for the decreased accuracy when the
energy splittings become larger. $B_0 = 1.9225\,{\rm cm}^{-1}$
is the rotational constant (Lovas et al. 1979), $l_{\rm sca} = 3$\,\AA\ is a
typical scattering length, $\mu = 3.5$\,amu is the reduced mass of the colliding
CO and He particles and $T$ is the kinetic gas temperature.

Finally, the rates for the inverse transitions were obtained from the condition
of detailed balancing, viz.

\begin{equation}
(2J^{\prime} + 1)\,\gamma_{J^{\prime}-J} =
(2J + 1)\,\gamma_{J-J^{\prime}}\,\exp{ (-h\,\nu_{J-J^{\prime}}/k\,T) }
\end{equation}

For $^{13}$CO, we used the same rate constants as for CO, with the proviso that
the mass difference potentially introduces an additional error on the collision
rates (3.5\%).

\end{document}